\documentclass[sn-mathphys,Numbered]{sn-jnl}% Math and Physical Sciences Reference Style
\usepackage{graphicx}%
\usepackage{multirow}%
\usepackage{amsmath,amssymb,amsfonts}%
\usepackage{amsthm}%
\usepackage{mathrsfs}%
\usepackage[title]{appendix}%
\usepackage{xcolor}%
\usepackage{textcomp}%
\usepackage{manyfoot}%
\usepackage{booktabs}%
\usepackage{algorithm}%
\usepackage{algorithmicx}%
\usepackage{algpseudocode}%
\usepackage{listings}%

\usepackage{physics}
\usepackage{verbatim}

\usepackage{csquotes}

\usepackage{soul}

\usepackage{bm}

\usepackage[normalem]{ulem}

\newenvironment{equations}
{\begin{equation}\begin{aligned}}
		{\end{aligned}\end{equation}\ignorespacesafterend}

\newcommand{\Schr}{Schr\"{o}dinger\ }
\newcommand{\dg}{\dagger}

\newcommand{\hL}{\hat{L}}
\newcommand{\hz}{\hat{z}}
\newcommand{\hB}{\hat{B}}
\newcommand{\hx}{\hat{x}}
\newcommand{\hp}{\hat{p}}

\newcommand{\prt}[1]{\left(#1\right)}
\newcommand{\prtq}[1]{\left[#1\right]}

\newcommand{\prtgB}[1]{\Bigg\{#1\Bigg\}}

\newcommand{\mcD}{\mathcal{D}}
\newcommand{\mcM}{\mathcal{M}}
\newcommand{\mcN}{\mathcal{N}}

\newcommand{\mcQ}{\mathcal{Q}}
\newcommand{\mcL}{\mathcal{L}}

\newcommand{\ad}{a^\dagger}

\newcommand{\intmp}{\int_{-\infty}^{+\infty}}

\newcommand{\hV}{\hat{V}}
\newcommand{\hM}{\hat{M}}

\newcommand{\tU}{\tilde{U}}

\begin{document}

\title[Article Title]{A proposal for a new kind of spontaneous collapse model}


\author*[1]{\fnm{Nicol\`{o}} \sur{Piccione}, \href{https://orcid.org/0000-0001-6391-2187}{Orcid}}\email{nicolo.piccione@neel.cnrs.fr}

\affil*[1]{\orgdiv{Institut N\'{e}el, Grenoble INP}, \orgname{Universit\'{e} Grenoble Alpes, CNRS}, \orgaddress{\street{25, Avenue des Martyrs}, \city{Grenoble}, \postcode{38000}, \country{France}}}


\abstract{Spontaneous collapse models are modifications of standard quantum mechanics in which a physical mechanism is responsible for the collapse of the wavefunction, thus providing a way to solve the so-called \enquote{measurement problem}. The two most famous of these models are the Ghirardi-Rimini-Weber (GRW) model and the Continuous Spontaneous Localisation (CSL) models.
Here, we propose a new kind of non-relativistic spontaneous collapse model based on the idea of collapse points situated at fixed spacetime coordinates. This model shares properties of both GRW and CSL models, while starting from different assumptions. 
We show that it can lead to a dynamics quite similar to that of the GRW model while also naturally solving the problem of indistinguishable particles. On the other hand, we can also obtain the same master equation of the 
CSL models. Then, we show how our proposed model solves the measurement problem in a manner conceptually similar to the GRW model. Finally, we show how the proposed model can also accommodate for Newtonian gravity by treating the collapses as gravitational sources.}

\keywords{Spontaneous Collapse Models, Measurement Problem, Indistinguishable particles, Quantum Foundations}



\maketitle

\section{Introduction\label{Sec:Introduction}}

Quantum mechanics is an extremely successful theory that arguably agrees with every direct experimental test conducted up to now. However, the principle of linear superposition apparently contradicts a commonplace observation: macroscopic objects are never found in a linear superposition of position states. Moreover, the theory does not explain why, during a quantum measurement, deterministic evolution is replaced by a probabilistic evolution whose random outcomes obey Born's probability rule. 

Spontaneous Collapse Models~\cite{Bassi2003Dynamical, Bassi2013Models,Bassi2023CollapseModels} offer a solution to this conundrum by modifying the \Schr dynamics. These modifications are negligible for a small quantum systems but become important when the mass or the number of particles of the system increase, leading to the collapse of the wavefunction in agreement with Born's rule. Moreover, the spontaneous collapse mechanism could be a key ingredient in explaining long-standing problems such as the existence of \enquote{dark energy} in cosmology~\cite{Josset2017DarkEnergy} and it has been proposed to be of gravitational origins~\cite{Karolyhazy1966Gravitation,Karolyhazy1986possible,Diosi1987Universal,Diosi1989Models,Penrose1996gravity,Bassi2017Gravitational,Gasbarri2017Gravity} or be the source of gravity~\cite{Tilloy2016CSLGravity,Tilloy2018GRWGravity}.

Since collapse models modify the \Schr dynamics, they make predictions which are different from those of standard quantum mechanics, particularly in the mesoscopic regime~\cite{Bassi2013Models}. These differences can be tested and, indeed, various experiments have been performed in order to test various kinds of spontaneous collapse models~\cite{Carlesso2019Collapse, Carlesso2022Present}. These experiments are not able to verify the actual dynamics that the quantum systems follow but they check for the modifications of the dynamics at the level of density matrix. Thus, different collapse models leading to the same density matrix would be experimentally equivalent.

Many kinds of spontaneous collapse models~\cite{Bassi2013Models} have been proposed, even very recently~\cite{Snoke2021Collapse}.
The two most famous families of spontaneous collapse models are the Ghirardi-Rimini-Weber (GRW) collapse models and the Continuous Spontaneous Localisation (CSL) models~\cite{Bassi2003Dynamical,Bassi2013Models,Bassi2023CollapseModels}. In GRW models, particles undergo spontaneous collapses happening at random points in space and time so that the evolution of a quantum system is unitary between two spontaneous collapses. On the other hand, in CSL models the \Schr equation is directly modified by the addition of new terms which make it stochastic and non-linear.

In this paper, we present a new type of spontaneous collapse model which we denote by \enquote{Collapse Points} (CP) model. The main idea consists of assuming that spacetime is permeated by collapse points inducing spontaneous weak measurements on quantum systems. This model is then conceptually different from the GRW and CSL models as the collapse dynamics is now something pertaining to the interaction between spacetime and quantum matter and not a property of quantum systems themselves. We derive a general stochastic equation obtained by coarse-graining over the collapse points and the related master equation. We show how the model can mimic the GRW model very well and how a natural assumption of the weak measurement operators leads to a model very similar to that of Refs.~\cite{Dove1995Symmetric,Tumulka2006spontaneous}, a GRW model consistent with indistinguishable particles. Then, we show how, with another choice of weak measurement operators, the CP model master equation can be exactly the same as the one of the CSL model, thus making our proposal compatible with experiments as the CSL model is. We then explain how the CP model solves the measurement problem in a similar way to the GRW and CSL models. Finally, we show how the collapses in the CP model can also be used to source Newtonian gravity in a similar way to the GRW model~\cite{Tilloy2018GRWGravity}, with the advantage that the CP model works also for indistinguishable particles and fields.

The paper is organised as follows. In Sec.~\ref{Sec:Model} we introduce the model and explain its properties. Then, in Sec.~\ref{Sec:StochasticEquation}, we coarse-grain over the collapse points to derive the stochastic \Schr equation and the related master equation. In Sec.~\ref{Sec:MeasurementProblem} we explain how the CP model solves the measurement problem. In Sec.~\ref{Sec:NewtonianGravity} we show how Newtonian gravity can be sourced by spontaneous collapses in the CP model. Finally, in Sec.~\ref{Sec:Conclusions}, we comment on the differences between our model and the GRW and CSL ones and briefly state the conclusions of our work.

\section{The collapse points model\label{Sec:Model}}

Here, we present the main elements of the new spontaneous collapse model we propose in this paper which we denote by \enquote{Collapse points} (CP) model. Despite the fact that the model presented here is a non-relativistic spontaneous collapse model, we motivate some of its assumptions by thinking of its relativistic extension, on which we plan to work in the future.

Let us consider an isolated quantum system $S$. In the model we propose, its state $\ket{\psi_t}$ evolves according to the system Hamiltonian but its dynamics is also disturbed by weak measurements taking place at fixed spacetime positions. These weak measurements can be modelled as interactions between the system and collapse points located at the spacetime positions where the weak measurements occur. Each of these interactions is then immediately followed by a spontaneous measurement of the collapse point, thus providing, in general, a weak measurement of the quantum system at hand. To be maximally consistent with Lorentz invariance, we assume that the collapse points are uniformly distributed in spacetime (i.e., according to a Poisson distribution in spacetime)\footnote{Such a distribution is Lorentz invariant in the sense that it appears the same in every frame.}. 

Our model is based on the idea of a flash ontology~\cite{Bell1987JumpsV1,Bell1987JumpsV2}\footnote{The word \enquote{flash} has been introduced by Tumulka in~\cite{Tumulka2006relativistic}.}, which is one of the ontologies proposed for the Ghirardi-Rimini-Weber (GRW) model. In GRW, a flash at point $x$ occurs when a particle undergoes a spontaneous collapse centred in $x$. Then, the purpose of the theory is to provide the probability distribution of such flashes~\cite{Esfeld2014GRW,Book_Tumulka2022Foundations}. In our model, for each collapse point undergoing a spontaneous measurement a flash can occur or not. It follows that the result of such measurement should be dichotomic, i.e., yes or no, \emph{flash} or \emph{no flash}. The simplest way to do this is to assume that at each collapse point we can associate a qubit initialised in state $\ket{0}$, the orthonormal one being denoted by $\ket{1}$. A flash occurs if such qubit is found in the state $\ket{1}$.

In the non-relativistic regime, we assume that the interaction between system $S$ and the collapse points can be taken to be instantaneous. While this could not be the case, even in principle, in the relativistic model because of possible incompatibilities with special relativity, we can motivate our assumption as follows. Taking the collapse radius associated to a weak measurement to be the same of the GRW model, i.e., $r_C \sim 10^{-7} {\rm m}$~\cite{Bassi2003Dynamical,Bassi2013Models}, we get that the duration $\tau_c$ of the interaction in the relativistic case should be of the following order of magnitude: $\tau_c \sim r_C/c \sim 10^{-16} {\rm s}$. This duration can be considered instantaneous for most relevant physical phenomena and this is why we make the assumption that the interaction can be considered instantaneous in the non-relativistic regime.

The interaction between system $S$ and a collapse point located at spacetime coordinates $\bar{x}_c = (x_c,t_c)$, where $x_c$ represent the spatial coordinates and $t_c$ the temporal one, is governed by the unitary operator $U_{x_c} = \exp{- i \sqrt{\gamma} H_{x_c}/\hbar}$, where $\sqrt{\gamma}$ can be seen as a parameter taking into account the effective duration of the interaction and/or the strength of the weak measurement. The interaction Hamiltonian $H_{x_c}$ is assumed, in generality, to be given by $H_{x_c} = \hL (x_c) \sigma_x$, where $\hL (x_c)$ is a generic Hermitian operator acting on the Hilbert space of system $S$ and $\sigma_x$, acting on the Hilbert space of the collapse point, is the usual Pauli operator such that $\sigma_x \ket{0} = \ket{1}$. It follows that, for a given system initial state at time $t=t_0$ and collapse points located at spacetime positions $(\bar{x}_1,\bar{x}_2,\dots,\bar{x}_n)$, with time coordinates increasing (i.e., $t_0 \leq t_1 \leq t_2 \leq \dots \leq t_n$), the joint probability distribution of flashes and no flashes is (see Appendix~\ref{APPSec:ProbabilityDistributionFlashes})
\begin{equations}
\label{eq:TotalExactEvolution}
P(\xi_1,\dots,\xi_n) &= \norm{\bra{\xi_1,\dots,\xi_n}U_n \dots U_1\ket{\psi_0,0_1,\dots,0_n}}^2,
\\
\ket{\psi^{(\xi_1,\dots,\xi_n)}} &= \frac{\bra{\xi_1,\dots,\xi_n}U_n \dots U_1\ket{\psi_0,0_1,\dots,0_n}}{\sqrt{P(\xi_1,\dots,\xi_n)}},
\end{equations}
where $\xi_m$ stands for either $1_m$ or $0_m$, i.e., flash or no flash at spacetime point $\bar{x}_m$. Morever, we defined $U_m = U_{x_m} U(t_m,t_{m-1})$, where $U(t_m,t_{m-1})$ is the evolution operator given by the Hamiltonian of system $S$ between times $t_{m-1}$ and $t_m$. The above equation shows that one can compute the flash probabilities by computing first the dynamics of the system wavefunction and collapse points without spontaneous measurements and then projecting on the desired flash/no flash distribution. Finally, we also notice that the probability of getting a flash or not at point $x_m$ can be computed as follows:
\begin{equation}
\label{eq:SingleFlashProbability}
P(\xi_m) = \Tr{U(t_m.t_{m-1})\rho_{m-1}U(t_m.t_{m-1})^\dg \otimes \dyad{\xi_m}},
\end{equation}
where the trace is over the joint Hilbert space of the quantum system and the $m$-th collapse point, and $\rho_{m-1}$ is the density matrix of the quantum system at time $t_{m-1}$ obtained by tracing out the collapse points up to $m-1$. Eq.~\eqref{eq:SingleFlashProbability} implies that the dynamics given by the CP model is Markovian.

% No faster-than-light communication because...

\section{Stochastic \Schr equation\label{Sec:StochasticEquation}}

We now consider that, at each collapse point, the probability of a collapse is extremely low. This condition has to be expected under normal circumstances because we have to recover the standard quantum dynamics for microscopic systems\footnote{In fact, with respect to the Ghirardi-Rimini-Weber (GRW) model, this corresponds to the fact that the probability of a spontaneous collapse for a single particle to occur, in the typical time-scale of laboratory observations, is extremely low. Analogously, with respect to the Continuous Spontaneous Localisation (CSL) models, this corresponds to the fact that a single quantum particle behaviour is basically the one predicted by standard quantum mechanics.}.
Moreover, we consider the density of collapse points to be high enough so that for lengths and times typically considered in quantum mechanics experiments there is a very high number of collapse points\footnote{A suggestive guess for the collapse points density could be that the spacetime volume containing on average one collapse point is given by the Planck spacetime volume.}. Under these assumptions, we can derive a stochastic \Schr equation which leads to a master equation potentially identical to that of the CSL model. This is done more in detail in Appendix~\ref{APPSec:DerivationStochasticEquation}. In the following we only report the main ingredients needed to obtain the stochastic equation.

Let us consider a system wavefunction $\ket{\psi_0}$ which is non-zero over a spatial volume $V$ and let us consider a time $\delta t$ which is short with respect to the system timescale. We denote by $N$ the average number of collapse points that is expected in the spacetime volume $c\delta t V$ and assume that $N \gg 1$. 
In other words, given a collapse points spacetime density $\mu$, we consider a large enough spacetime volume $c\delta t V$ such that $N=\mu c \delta t V \gg 1$.
Since the probability of a flash at each collapse point is extremely low we can compute the probability of no flashes at first order in $\gamma$ and obtain
\begin{equation}
P(0) \simeq 1 - \frac{\gamma}{\hbar^2} \sum_{j=1}^{\lfloor N \rfloor} \ev{\hL^2 (x_j)}{\psi_0},
\end{equation}
where we neglected the bare evolution of the system because of the short timescale $\delta t$ and $x_j$ denotes the spatial position of the $j$-th collapse point.
We can also compute, again at first order in $\gamma$, the probability of getting a flash at the collapse point located in $x_j$, obtaining
\begin{equation}
P(x_j) \simeq \frac{\gamma}{\hbar^2}\ev{\hL^2 (x_j)}{\psi_0},
\end{equation}
where we observe that, at this order of perturbation, $P(0) + \sum_j P(x_j) = 1$, which means that, at this order, the probability of getting two flashes must be neglected.

The above probabilities can be recast in a continuous form by making a coarse-grain over the collapse points. By making use of the previously introduced collapse points density $\mu = N/(c \delta t V)$ and exploiting the uniform distribution of the collapse points in the spacetime volume $c \delta t V$, we can write the probability density of getting a flash around point $x$ as:
\begin{equation}
\label{eq:FlashProbability}
P(x\in \dd[3]{x})\dd[3]{x} \simeq \delta t\frac{c \gamma \mu}{\hbar^2} \dd[3]{x} \ev{\hL^2 (x)}{\psi_0},
\end{equation}
while the probability of no-flash is given by
\begin{equation}
\label{eq:NoFlashProbability}
P(0) \simeq 1 - \delta t\frac{c\gamma \mu}{\hbar^2} \int \dd[3]{x} \ev{\hL^2 (x)}{\psi_0}.
\end{equation}
Considering the effect that the occurrence of a flash has on the wavefunction and restoring its Hamiltonian evolution we finally arrive at the stochastic \Schr equation:
\begin{multline}
\label{eq:StochasticSchrodingerEquation}
\dd{\ket{\psi(t)}}
= \Bigg[-\frac{i}{\hbar} H \dd{t} + 
\frac{c \gamma \mu}{2\hbar^2}\int \dd[3]{x} \dd{t} \Big(\ev{\hL^2 (x)}\\-\hL^2(x)\Big)
- \int \dd[3]{x} \frac{\dd{N_x}}{\dd[3]{x}}\Bigg(1 + \frac{i \hL(x)}{\sqrt{\ev{\hL^2 (x)}}}\Bigg)\Bigg] \ket{\psi (t)}
\end{multline}
where we have the Poisson process $\dd{N_x}$ such that $\mathbb{E}\prtq{\dd{N_x}}=P(x\in \dd[3]{x})\dd[3]{x}$.

The dynamics described by Eq.~\eqref{eq:StochasticSchrodingerEquation} can be strikingly similar to that of the GRW model if applied to a single particle. For example, if we choose $\hL (x)$ as in GRW~\cite{Bassi2013Models}, for a single particle we get that $\int \dd[3]{x} \ev{\hL^2 (x)}{\psi_0} = 1$ and, therefore, $(c\gamma \mu)/\hbar^2$ can be interpreted as the collapse rate $\lambda_{\rm GRW}$ of GRW [cf. Eqs.~\eqref{eq:FlashProbability} and~\eqref{eq:NoFlashProbability}]. Then, the probability of getting a flash at point $x$ is the same as in GRW [cf. Eq.~\eqref{eq:FlashProbability}] and, as in GRW, the wavefunction changes as follows
\begin{equation}
\ket{\psi (t)} \xrightarrow{\textrm{Flash at $x$}} \frac{\hL (x) \ket{\psi (t)}}{\sqrt{\ev{\hL^2 (x)}{\psi (t)}}}.
\end{equation}
The only difference with GRW is that the particle wavefunction is slightly modified even in absence of flashes.

We can also easily re-derive a version of GRW which is compatible with indistinguishable particles~\cite{Dove1995Symmetric,Tumulka2006spontaneous,Book_Tumulka2022Foundations}. We can make the natural assumption that the probability of a flash has to be proportional to the number of particles at or around the collapse point. Even better, we can assume that the probability of a flash has to be proportional to the mass as is usually done for the collapse operators in CSL models~\cite{Bassi2013Models}. We then write
\begin{gather}
\hL (x) = \sqrt{\mcM (x)},
\quad
\mcM (x) = \sum_{i} \frac{m_i}{m_R} \mcN_i (x), \nonumber
\\
\mcN_i (x) = \intmp \dd{y} g(y-x) \ad_i(y) a_i(y),
\label{eq:MassOperator}
\end{gather}
where the subscript $i$ indicates the $i$-th kind of particle considered in the model,
$a_i (x)$ and $\ad_i (x)$ are, respectively, annihilation and creation operators of the $i$-th kind of particle at the spatial point $x$, and $g(x)$ is a smearing function which is usually a Gaussian centered on $x$~\cite{Dove1995Symmetric,Bassi2003Dynamical,Bassi2013Models,Book_Tumulka2022Foundations}. Then, $\mcN (x)$ is a smeared number operator at point $x$ for the $i$-th kind of particles and $\mcM (x)$ is a smeared mass operator obtained with weights $m_i/m_R$ where $m_i$ is the mass of the $i$-th kind of particle and $m_R$ is a reference mass, usually taken as that of a nucleon. By making these assumption in the language of quantum field theory, we can easily re-obtain the collapse operators proposed in Ref.~\cite{Dove1995Symmetric}. This is shown in detail in Appendix~\ref{APPSec:IndistinguishableParticles}. In Sec.~\ref{Sec:Conclusions} we will discuss why the mass operator can be argued to be a natural choice for $\hL^2 (x)$.

Regarding the comparison with CSL, Eq.~\eqref{eq:StochasticSchrodingerEquation} is quite different from the stochastic \Schr equation characterizing the CSL model. In fact, while the latter contains gaussian noise Eq.~\eqref{eq:StochasticSchrodingerEquation} presents the occurence of Poisson noise. This means that in the CSL model the wavefunction reduction takes place gradually while in our model it takes place abruptly, like in the GRW model. However, in Appendix~\ref{APPSec:MasterEquationDerivation}, we show that both our model and the CSL model can lead to the same master equation, thus making all of the experiments done with regard to the CSL model also valid for our model. We report here the result:
\begin{equation}
\label{eq:MasterEquation}
\dv{t} \rho_t = - \frac{i}{\hbar} \comm{H}{\rho_t} + \frac{c\mu\gamma}{\hbar^2} \int \dd[3]{x} \mcD_{\hL (x)} (\rho_t),
\end{equation}
where $\mcD_{A}$ stands for the standard Lindbladian dissipator associated to the jump operator $A$:
\begin{equation}
	\mcD_{A} (\rho_t) = A \rho_t A^\dg - \frac{1}{2}\acomm{A^\dg A}{\rho_t}.
\end{equation}
By appropriately choosing the operator $\hL (x)$ we can make Eq.~\eqref{eq:MasterEquation} exactly the same as the one of the CSL model~\cite{Bassi2003Dynamical,Bassi2013Models}.

\section{How the CP model solves the measurement problem\label{Sec:MeasurementProblem}}

As the other spontaneous collapse models, the CP (Collapse Points) model aims to solve the famous measurement problem. Conceptually, the way the problem is solved is the same of the other collapse models: we modify the standard quantum dynamics by introducing a new dynamical element such as spontaneous collapses in GRW, noise terms in CSL, and collapse points in the CP model. This new dynamical element is practically negligible for small quantum systems but becomes more and more important as the system approaches macroscopicity (either by number of particles or by its mass). This is usually called the amplification mechanism~\cite{Bassi2003Dynamical,Bassi2013Models}. We have now to argue why this would bring to the Born rule in the CP model and the general dynamics associated with a textbook projective measurement.

Following the von Neumann measurement scheme~\cite{Book_Wiseman2009Measurement}, we consider the following measurement dynamics:
\begin{equation}
\dyad{\phi_0} \otimes \rho_0^{(A)} \rightarrow \sum_{i,j} c_i c_j^* \dyad{\lambda_i}{\lambda_j} \otimes \rho_{i,j}^{(A)}\end{equation}
where $\ket{\phi_0} = \sum_i c_i \ket{\lambda_i}$ is the initial state of the system we want to measure, $\ket{\lambda_i}$ are the eigenstates of the observable we want to measure\footnote{We assumed the observable to not have nondegenerate eigenvalues for simplicity.}, $\rho_0^{(A)}$ is the initial state of the apparatus used for the measurement and the $\rho_{i,j}^{(A)}$ are such that $\rho_{i,i}^{(A)}$ and $\rho_{j,j}^{(A)}$ with $i\neq j$ represent the apparatus occupying macroscopically different regions of space while the matrices $\rho_{i,j}^{(A)}$ represent the quantum coherences between these different regions of space. Eq.~\eqref{eq:SingleFlashProbability} shows that the probability of getting a flash or not at certain point can be obtained by knowing the density matrix of the system right before its interaction with the collapse point. In the non-relativistic limit the density matrix of a system evolves according to Eq.~\eqref{eq:MasterEquation}, which, when choosing the same collapse operators as in CSL, leads to decoherence in position basis~\cite{Bassi2003Dynamical,Bassi2013Models}. Then, the final state $\rho_f$ obtained by tracing out the collapse points is 
\begin{equation}
\rho_f \simeq \sum_i \abs{c_i}^2 \dyad{\lambda_i} \otimes \rho_{i,i}^{(A)},
\end{equation}
where we remark that to get the above equation we considered all modifications due to the dissipator in Eq.~\eqref{eq:MasterEquation} to be negligible apart from the suppression of spatial decoherence. At this stage, considering the macroscopicity of the apparatus, the probability of getting at least one flash or at least, say, one-thousand flashes is practically one. However, the probability of getting two flashes corresponding to different $\lambda_i$ is practically zero because a flash indicates where the macroscopic object is\footnote{Choosing the same operators as in CSL we encounter the same \enquote{tail problem} as in GRW and CSL which, however, does not pose any real practical problem.}. It follows that the probability of getting flashes in the macroscopic area identified by $\rho_{i,i}^{A}$ is $\abs{c_i}^2$, i.e., it follows the Born rule. Moreover, once we see at least one flash corresponding to such an area, say for $\rho_{n,n}^{(A)}$ the wavefunction of system and apparatus is now known to be $\rho_f = \dyad{\lambda_n}\otimes \rho_{n,n}^{(A)}$ so that we also explain the reduction postulate.

\section{Collapse points as Newtonian gravity sources\label{Sec:NewtonianGravity}}

In this section, inspired by the work of Tilloy in Ref.~\cite{Tilloy2018GRWGravity}, we show how the collapse points can consistently act as gravitational sources for a Newtonian classical gravitational field. With respect to Ref.~\cite{Tilloy2018GRWGravity}, this construction has the advantage of naturally dealing with indistinguishable particles and fields, despite being still based on flashes as the source of the gravitational field. This section purpose is to show an example of how the CP model can lead to more sensible results than the standard GRW model, while still being based on a flash ontology. A full characterisation of a Newtonian gravity model based on these premises, like the one done in Ref.~\cite{Tilloy2016CSLGravity}, is beyond the scope of the present paper.

Let us associate to each collapse point the following Poisson equation for the gravitational potential:
\begin{equation}
	\label{eq:PoissonEquation}
	\Delta \Phi_j (x,t,\xi_j) = 4 \pi G m_R \lambda_{\rm GRW}^{-1} f_j (x,t,\xi_j),
\end{equation}
where $\Phi_j (x,t)$ is the gravitational potential associated to the $j$-th collapse point situated in $(x_j,t_j)$, $m_R$ is the reference mass used to define $\mcM(x)$ [cf. Eq.\eqref{eq:MassOperator}], $\lambda_{\rm GRW} \equiv (c\gamma\mu)/\hbar^2$, and $f_j (x,t,\xi_j)$ is a spacetime form factor that smears the effect of the flash in spacetime, as in Ref.~\cite{Tilloy2018GRWGravity}. It satisfies $\int \dd[3]{x} \dd{t} f_j (x,t,1) = 1$, $f\geq 0$, $t<t_j \implies f_j (x,t,\xi_j) = 0$, and $f(x,t,0)=0$\footnote{We recall that $\xi_j$ represents the occurrence of a flash at the collapse points located in $(t_j,x_j)$; $\xi_j$=0 means that no flash occurred and $\xi_j=1$ means that the flash occurred.}. With respect to the definitions given in Ref.~\cite{Tilloy2018GRWGravity}, we have to assume that each flash generates the same kind of gravitational field because the flashes are agnostic to what kind of particles or field caused them. However, making the natural choice\footnote{We argue why this choice is natural in sec.~\ref{Sec:Conclusions}.} $\hL (x) = \sqrt{\mcM(x)}$ [see Eq.~\eqref{eq:MassOperator}], the number of flashes will be proportional to the amount of mass, allowing for the correct recovery of the Newtonian limit.

The formal solution of Eq.~\eqref{eq:PoissonEquation} is
\begin{equation}
\label{eq:GravitationalPotential}
\Phi_j (x,t,\xi_j) = - G m_R \lambda_{\rm GRW}^{-1} \int \dd[3]{y} \frac{f_j (y,t,\xi_j)}{\abs{x-y}},
\end{equation}
and the effect of the gravitational potential generated by a collapse point on quantum systems is given by the external potential
\begin{equation}
	\hV_j^{(G)} (t) = \int \dd[3]x \Phi_j (x,t,\xi_j) \hM (x),
\end{equation}
where $\hM(x)$ is the mass operator, equal to the smeared one defined in Eq.~\eqref{eq:MassOperator} with $g(y-x)=\delta(y-x)$ and multiplied by $m_R$.
The simplest choice for the functions $f_j$ is $f_j (x,t,\xi_j) = \delta(x-x_j)\delta(t-t_j)\delta_{\xi_j,1}$, which, however, leads to a divergent energy variation (see Appendix~\ref{APPSec:NewtGravEnergy}). Therefore, we make the second obvious choice, which is $f_j (x,t,\xi_j) = \delta(t-t_j)\delta_{\xi_j,1}f_G (\abs{x-x_j})$ where $f_G (\abs{x-x_j})$ is a smearing function in space which only depends on the distance from the $j$-th collapse point. This choice makes the potential $\hV_j^{(G)} (t=t_j)$ divergent in time but this can be accommodated inside the related evolution operator $\tU_G (x_j) = \exp[-(i/\hbar) \int \hV^{(G)}_j (t)\dd{t}]$, which in this limit gives
\begin{equation}
	\tU_G (x_j) = e^{i \frac{G m_R}{\lambda_{\rm GRW}\hbar} \delta_{\xi_j,1} \int \dd[3]{x} \dd[3]{y} \frac{f_G(\abs{y-x_j})\hM (x)}{\abs{x-y}}}.
\end{equation}

At the level of the coarse-grained dynamics of Eq.~\eqref{eq:StochasticSchrodingerEquation}, the inclusion of such gravitational effect corresponds to defining new collapse operators $\hB (x) \equiv U_G (x) \hL (x)$, where $U_G (x) \equiv \tU_G (x)$ with $\delta_{\epsilon_x,1} \rightarrow 1$. The related stochastic \Schr equation and master equation can be immediately obtained by making the substitution $\hL (x) \rightarrow \hB (x)$ in Eqs.~\eqref{eq:StochasticSchrodingerEquation} and~\eqref{eq:MasterEquation}\footnote{Notice, however, that $\hL^2 \rightarrow \hL^2$ in the stochastic equation because this term is tied to the non-occurrence of flashes.}. We report here the new master equation for commodity of the reader:
\begin{equation}
\label{eq:GravitationalMasterEquation}
\dv{t} \rho_t = - \frac{i}{\hbar} \comm{H}{\rho_t} + \lambda_{\rm GRW}\int \dd[3]{x} \mcD_{\hB (x)} (\rho_t),
\end{equation}
which is indeed different from the one obtained in Ref.~\cite{Tilloy2018GRWGravity} because the wavefunction gets modified even when flashes do not occur.

On the macroscopic scale, the Newtonian potential $\Phi_G (x,t)$ can be recovered as follows. We consider the average potential coming by summing the potential due to all flashes
\begin{equation}
\label{eq:TotalAveragePotential}
\Phi_G (x,t) = \int \dd{s}\dd[3]{y} \Phi (x,t,y,s) P_f (y,s),
\end{equation}
where $\Phi (x,t,y,s)$ is the potential at point $(x,t)$ generated by a flash at spatial point $y$ and time $s$ and $P_f (y,s) \dd{s} \dd[3]{y}$ is the mean number of flashes in the infinitesimal spacetime volume around spatial point $y$ and time $s$\footnote{Equivalently, $P_f (y,s)$ is the
probability density of getting a flash at point $y$ and time $s$.}. Now, we simply make the following substitutions [cf. Eq.~\eqref{eq:FlashProbability} and Eq.~\eqref{eq:GravitationalPotential}]:
\begin{equations}
\Phi (x,t,y,s) &= - G m_R \lambda^{-1}_{\rm GRW} \delta(t-s) \int \dd[3]{z} \frac{f_G (\abs{y-z})}{\abs{x-z}},
\\
P_f (y,s) &= \lambda_{\rm GRW} \ev{\mcM (y)}_s,
\end{equations}
where we choose $\hL(x)=\sqrt{\mcM(x)}$ for the collapse operator, $f(y,t,z,s)=\delta(t-s)f_G(\abs{y-z})$, and we denoted by $\ev{\mcM(x)}_s$ the average mass distribution at time $s$. Substituting this into Eq.~\eqref{eq:TotalAveragePotential} we get
\begin{equation}
\Phi_G (x,t) = - G \int \dd[3]{y} \dd[3]{z} \frac{m_R f_G(\abs{y-z})\ev{\mcM(y)}_t}{\abs{x-y}}.
\end{equation}
The function $f_G (x)$ is characterised by a radius $r_G$, which we can choose to be similar to the collapse radius $r_C \sim 10^{-7} {\rm m}$. On a scale much larger than this, $f_G (\abs{y-z}) \sim \delta (y-z)$ and we get
\begin{equation}
\Phi_G (x,t) \simeq - G \int \dd[3]{y} \frac{m_R \ev{\mcM(y)}_t}{\abs{x-y}},
\end{equation}
which is the Newtonian potential generated by the mass distribution $m_R \ev{\mcM(y)}_t$ at time $t$\footnote{Notice that $m_R$ appears in the denominator of the smeared mass operator [cf. Eq.~\eqref{eq:MassOperator}].}.

In Appendix~\ref{APPSec:NewtGravEnergy}, we briefly study the dynamics due to Eq.~\eqref{eq:GravitationalMasterEquation} on a single particle, neglecting the particle bare Hamiltonian. As in Ref.~\cite{Tilloy2018GRWGravity}, we obtain that the sole effect is a decay of the off-diagonal elements of the particle density matrix in position basis, enhanced by gravity. In particular, using the GRW collapse operators and the simplest choice $f_G(\abs{x-z})=\delta(x-z)$,
the dephasing $\Gamma(d)$ behaves as $\Gamma(d) \propto d^{3/2}/\sqrt{\lambda_{\rm GRW}}$ for $d \ll r_m^3/r_C^2, r_C^2/r_m$, where $d$ is the distance between the two considered spatial points and $r_m \equiv G m_R m/(\hbar \lambda_{\rm GRW})$.
Therefore, as in Ref.~\cite{Tilloy2018GRWGravity}, this calculation hints that the inclusion of gravity in this way leads to the possibility, in principle, of experimental bounds from below for $\lambda_{\rm GRW}$.

Overall, we showed that the CP model can accommodate a Newtonian gravity sourced by flashes in the same way as GRW can~\cite{Tilloy2018GRWGravity}. However, contrarily to the GRW case, this construction works for indistinguishable particles, particle creation and destruction, and fields. A full characterisation of such a model will be the subject of a future work.

\section{Discussions and conclusions\label{Sec:Conclusions}}

In this paper we presented the idea for a new kind of spontaneous collapse model in which spacetime is permeated by collapse points responsible for spontaneous weak measurements performed on wavefunctions. Dynamically, the collapse points (CP) model is located half-way between the GRW and CSL models, the two most discussed spontaneous collapse models. It presents the same abrupt jumps of GRW models but the dynamics of the wavefunction is slightly modified even when no flashes occur. Conceptually, however, the CP model presents some differences with the other two. In GRW models, the collapse is something pertinent to the particles while in CSL models it is something pertinent to the fields. In the CP model, collapses do not belong to a specific particle or field but to spacetime itself. It could be the case then that this model, when developed, could be more friendly to be justified for gravitational reasons or even be compatible with general relativity. In fact, gravity has already been proposed in various ways as the source of spontaneous collapses in the literature~\cite{Karolyhazy1966Gravitation,Karolyhazy1986possible,Diosi1987Universal,Diosi1989Models,Penrose1996gravity,Bassi2017Gravitational,Gasbarri2017Gravity}. Moreover, such kind of explanations would be well-suited for that line of research trying to solve the problem of a unified theory of quantum mechanics and gravity by keeping the spacetime classical and not making it quantum~\cite{Penrose2014Gravitization,Bassi2017Gravitational,Oppenheim2021postquantum,Oppenheim2022Constraints}. 

In light of the previous discussion, it seems to us natural that one would like the probability of a flash to depend on the amount of \enquote{content} at or around each collapse point. The most natural way to account for this would involve the energy-momentum tensor, also in light of Einstein's field equations. In the non-relativistic limit, this leads to considering the amount of energy first, and, then, to neglect all terms apart from $m c^2$, i.e., the amount of mass. This is why we consider the amount of mass as a natural choice for $\hL^2 (x)$ in Eq.~\eqref{eq:MassOperator} of Sec.~\ref{Sec:StochasticEquation}.

For the GRW and CSL models, many extension to the relativistic regime have been proposed~\cite{Pearle1990Toward, Dove1996Local, Dove1996Explicit, Breuer1998Relativistic, Breuer1999stochastic, Pearle1999Relativistic, Nicrosini2003Relativistic, Dowker2004SpontaneousLattice, Tumulka2006relativistic, Bedingham2011Relativistic, Bedingham2014Matter, Tilloy2017interacting, Bedingham2019csl, Tumulka2021relativistic}, despite the difficulty of such extension~\cite{Jones2021Impossibility,Jones2021MassCoupled,Thesis_Jones2021Special}. We argue that also the CP model can be extended to the relativistic domain. First, because relativistic extensions of the GRW and CSL models already exist. Second, because the works of Aharanov and Albert~\cite{Aharonov1980States,Aharonov1981RelativisticMeasurement,Aharonov1984TimeNonRelativistic,Aharonov1984TimeRelativistic,Book_Tumulka2022Foundations} show how standard quantum measurement, localised at certain spacetime points, can be made compatible with special relativity. In the CP model, measurements take place in pre-determined spacetime points so that this extension seems quite possible even for this second reason. However, while measurements in the CP model are ideal as the measurements considered in this literature, the interaction that we considered between a collapse point and the wavefunction is not local, at least in the non-relativistic CP model presented in this work. Therefore, the extension of the CP model to the relativistic regime is not automatic and a lot of effort will be required to do it.

In this paper, we presented the Collapse Points model and the way it appears in its non-relativistic version.
We showed how the model can mimic the GRW model and naturally assume the form that solves the problem of indistinguishable particles and the application to fields.
We also showed that making certain choices the model seems to behave (at the density matrix level) like the CSL model. In particular, making a certain choice for the interaction operators between a quantum system and the collapse points we were able to re-derive the CSL master equation, thus making the CP model as experimentally tested as CSL. Then, we explained how the CP model solves the measurement problem in a similar way to the other collapse models. Finally, we showed how the collapse points can be used to source Newtonian gravity. This was already been done for the GRW model~\cite{Tilloy2018GRWGravity}, but with the CP model this gravitational model also applies to indistinguishable particles and fields.

\section*{Acknowledgements}

This work was supported by the John Templeton Foundation (Grant No. 61835). We thank A. Bassi, and R. Tumulka for useful discussions about the premises of this work, A. Bassi, S. Donadi, J.L. Gaona-Reyes, and A. Tilloy for technical discussions, and M. Paternostro for his feedback on it.

\section*{Declarations}

\begin{itemize}
	\item Funding: This work was supported by the John Templeton Foundation (Grant No. 61835).
	\item Conflict of interest/Competing interests: Not applicable
	\item Ethics approval: Not applicable
	\item Consent to participate: Not applicable
	\item Consent for publication: Not applicable
	\item Availability of data and materials: Not applicable
	\item Code availability: Not applicable
	\item Authors' contributions: There is only one author
\end{itemize}

\noindent
If any of the sections are not relevant to your manuscript, please include the heading and write `Not applicable' for that section.

\clearpage

\begin{appendices}

\section{Probability distribution of flashes\label{APPSec:ProbabilityDistributionFlashes}}

In this Appendix, we give more details needed to prove the validity of Eq.~\eqref{eq:TotalExactEvolution}.

Let us suppose that, at time $t=t_0$, the state of system $S$ is given by $\ket{\psi_0}$. Up to the first collapse point the wavefunction evolves according to $U_1 = U_{x_1}U(t_1-t_0)$ where $U(t_1-t_0)$ is the evolution operator given by the Hamiltonian of system $S$ and $U_{x_1}$ is the instanteous interaction with the collapse point located in $x_1$ at time $t_1$. The probability of getting a flash (or not) in $x_1$ and the subsequent wavefunction are given by
\begin{equation}
P(\xi_1) = \norm{\mel{\xi_1}{U_1}{\psi_0,0_1}}^2,
\quad
\ket{\psi^{(\xi_1)}} = \frac{\mel{\xi_1}{U_1}{\psi_0,0_1}}{\sqrt{P(\xi_1)}},
\end{equation}
where $\xi$ stands for either $0$ or $1$ or, respectively, no flash and flash. Then, we have another spontaneous measurement with the second collapse point located in $x_2$. In this case, we have
\begin{multline}
P(\xi_2|\xi_1) 
= \norm{\mel{\xi_2}{U_2}{\psi^{(\xi_1)},0_2}}^2
= \norm{\bra{\xi_2}U_2\frac{\mel{\xi_1}{U_1}{\psi_0,0_1}}{\sqrt{P(\xi_1)}}\ket{0_2}}^2
=\\= \frac{1}{P(\xi_1)} \norm{\bra{\xi_1,\xi_2}U_2 U_1\ket{\psi_0,0_1,0_2}}^2,
\end{multline}
where $U_2 = U_{x_2}U(t_2-t_1)$ and which implies that 
\begin{equation}
P(\xi_1,\xi_2) = P(\xi_2 | \xi_1) P(\xi_1) = \norm{\bra{\xi_1,\xi_2}U_2 U_1\ket{\psi_0,0_1,0_2}}^2.
\end{equation}
The state is then given by
\begin{equation}
\ket{\psi^{(\xi_1,\xi_2)}} 
= \frac{\bra{\xi_2} U_2 \ket{\psi^{(\xi_1)},0_2}}{\sqrt{P(\xi_2|\xi_1)}}
= \frac{\bra{\xi_1,\xi_2}U_2U_1\ket{\psi_0,0_1,0_2}}{\sqrt{P(\xi_1,\xi_2)}}.
\end{equation}
By induction, it follows that
\begin{equations}
\label{APPeq:TotalExactEvolution}
P(\xi_1,\dots,\xi_n) &= \norm{\bra{\xi_1,\dots,\xi_n}U_n \dots U_1\ket{\psi_0,0_1,\dots,0_n}}^2,
\\
\ket{\psi^{(\xi_1,\dots,\xi_n)}} &= \frac{\bra{\xi_1,\dots,\xi_n}U_n \dots U_1\ket{\psi_0,0_1,\dots,0_n}}{\sqrt{P(\xi_1,\dots,\xi_n)}},
\end{equations}
which is Eq.~\eqref{eq:TotalExactEvolution} of the main text.

\clearpage
\section{Derivation of the stochastic \Schr equation\label{APPSec:DerivationStochasticEquation}}

The aim of this Appendix is to derive the stochastic \Schr equation given in the main text [Eq.~\eqref{eq:StochasticSchrodingerEquation}].

We consider the system wavefunction to be confined in a volume $V$ and study its evolution for a time $\delta t$ which is very short with respect to the time scale of $H$, so that we can ignore it. First, we consider the joint dynamics of wavefunction and collapse points up to first order in $\gamma$. We assume that in the spacetime volume $c\delta t V$ there are, on average, $N$ collapse points, where $N\gg 1$. In fact, the average number of collapse points in a given spacetime region is not a natural number but since it is much higher than one, we can round it down without harm, so we treat $N$ as a natural number.
Starting from the unitaries in Eq.~\eqref{eq:TotalExactEvolution} we get:
\begin{multline}
U_N \dots U_1 \simeq \prod_{j=1}^N \prtq{1 -i \frac{\sqrt{\gamma}}{\hbar}\hL \prt{x_j}\sigma_x^{(j)} - \frac{\gamma}{2 \hbar^2} \hL^2 \prt{x_j}}
\simeq\\ \simeq 
1 -i \frac{\sqrt{\gamma}}{\hbar}\sum_{j=1}^N \hL \prt{x_j} \sigma_x^{(x_j)} 
-\frac{\gamma}{2 \hbar^2} \sum_{j=1}^N \hL^2 (x_j)
- \frac{\gamma}{\hbar^2} \sum_{k>j}\hL (x_k)\hL (x_j)\sigma_x^{(x_k)}\sigma_x^{(x_j)}.
\end{multline}
Then, the probability of getting no flashes is
\begin{multline}
P(0) \equiv  P(0_1,\dots,0_n) = \norm{\bra{0_1,\dots,0_n}U_n \dots U_1\ket{\psi_0,0_1,\dots,0_n}}^2 
\simeq \\ \simeq 1 - \frac{\gamma}{\hbar^2} \sum_{j=1}^N \ev{\hL^2 (x_j)}{\psi_0},
\end{multline}
while the probability of getting a flash at point $x_j$ is given by
\begin{equation}
P(x_j) \equiv P(0_1,\dots,1_j,\dots,0_N) \simeq \frac{\gamma}{\hbar^2}\ev{\hL^2 (x_j)}{\psi_0}.
\end{equation}
We can see that at this order of perturbation $P(0) + \sum_j P(x_j) = 1$, which means that, at this order, the probability of getting two flashes must be neglected. The wavefunction in the two cases evolves as follows:
\begin{equations}
\label{APPeq:SingleCollapseEvolution}
\ket{\psi^{(0)}} &= \prtq{1 + \frac{\gamma}{2\hbar^2}\sum_{j=1}^N \prt{\ev{\hL^2 (x_j)}{\psi_0}-\hL^2 (x_j)}}\ket{\psi_0},
\\
\ket{\psi^{(x_j)}} &= \frac{-i \hL (x_j)}{\sqrt{\ev{\hL^2 (x_j)}{\psi_0}}}\ket{\psi_0}.
\end{equations}

Now, we consider the fact that the collapse points are uniformly distributed in the spacetime volume $c \delta t V$. We define $\mu = N/(c \delta t V)$ as the collapse point density in spacetime and we compute the probability of flash or no flash as
\begin{equations}
P(0) 
&\simeq 1 - \frac{\gamma N}{\hbar^2 V} \int_V \dd[3]{x} \ev{\hL^2 (x)}{\psi_0}
\simeq 1 - \delta t\frac{c\gamma \mu}{\hbar^2 } \int \dd[3]{x} \ev{\hL^2 (x)}{\psi_0},
\\
P(1) 
&\simeq \frac{\gamma N}{\hbar^2 V} \int_V \dd[3]{x} \ev{\hL^2 (x)}{\psi_0}
\simeq \delta t\frac{c \gamma \mu}{\hbar^2} \int \dd[3]{x} \ev{\hL^2 (x)}{\psi_0}.
\end{equations}
The probability of getting a flash around point $x$ is 
\begin{equation}
P(x\in \dd[3]{x})\dd[3]{x} \simeq \delta t\frac{c \gamma \mu}{\hbar^2} \dd[3]{x} \ev{\hL^2 (x)}{\psi_0}
\end{equation}
Then, when there is no flash, or there is a flash at point $x$, the state changes according to
\begin{equations}
\ket{\psi^{(0)}} &= \prtq{1 - \delta t\frac{c \gamma \mu}{2\hbar^2}\int \dd[3]{x} \prt{\hL^2 (x)-\ev{\hL^2 (x)}{\psi_0}}}\ket{\psi_0},
\\
\ket{\psi^{(x)}} &= \frac{-i \hL (x)}{\sqrt{\ev{\hL^2 (x)}{\psi_0}}}\ket{\psi_0}.
\end{equations}

Following Refs.~\cite{Jacobs2006straightforward,Book_Wiseman2009Measurement} the above leads to the stochastic equation
\begin{equation}
\label{APPeq:StochasticSchrodingerPoisson}
\dd{\ket{\psi(t)}}
=
\int \dd[3]{x} \prtq{\dd{t}\frac{c \gamma \mu}{2\hbar^2} \prt{\ev{\hL^2 (x)}-\hL^2(x)} - \frac{\dd{N_x}}{\dd[3]{x}}\prt{1 + \frac{i \hL(x)}{\sqrt{\ev{\hL^2 (x)}}}}} \ket{\psi (t)},
\end{equation}
where we have the Poisson process $\dd{N_x}$ such that $\mathbb{E}\prtq{\dd{N_x}}=P(x\in \dd[3]{x})\dd[3]{x}$ (with $\delta t \rightarrow \dd{t}$). The Poisson process has the property $\dd{N_x}^2 = \dd{N_x}$ because its result can only be zero or one. Moreover, one has that $\dd{N_x}\dd{t} = 0$ and $\dd{N_x}\dd{N_y} = \delta (x-y)\dd[3]{y}\dd{N_x}$. Eq.~\eqref{APPeq:StochasticSchrodingerPoisson} can be seen as the spatially continuous version of Eq. (3.156) of Ref.~\cite{Book_Jacobs2014MeasurementTheory}, with the identification $a \rightarrow -i L$. Restoring the evolution due to the Hamiltonian $H$ of the system we get Eq.~\eqref{eq:StochasticSchrodingerEquation} of the main text.

\clearpage
\section{Collapse operators for indistinguishable particles\label{APPSec:IndistinguishableParticles}}

In this section we show how to re-obtain the collapse operator defined in Ref.~\cite{Dove1995Symmetric} by starting from the smeared mass operator of Eq.~\eqref{eq:MassOperator}. We recall that we consider the coarse-grain operation of the collapse points and the expansions at first order in $\gamma$ already effectuated. We start with a bosonic field and then explain why the same calculation holds for a fermionic field. The sum of the smeared number operators preceded by the factors $m_i/m_R$ gives the smeared mass operator and its application to wavefunctions with a precise number of particles for each kind of particle gives the same collapse operator defined in Ref.~\cite{Dove1995Symmetric}. Indeed, if we consider a simplified toy model with just one kind of particles, we do not need the smeared mass operator and we can just use the smeared number operator.

We consider a single type of indistinguishable particles. We want the probability of a flash around point $x$ to be proportional to the amount of particles so that [cf. Eq.~\eqref{eq:FlashProbability}] we require $\hL^2 (x) \propto \mcN (x)$, where $\mcN (x)$ is the smeared number operator defined as follows:
\begin{equation}
\mcN (x) \equiv \intmp \dd{y} g(y-x) \ad (y) a (y),
\end{equation}
where $a (x)$ and $\ad(x)$ are, respectively, annihilation and creation operators at the spatial point $x$. The action of such operator on a state $\ket{\psi_n}$ with a fixed number $n$ of particles is the following:
\begin{equations}
\mcN (x) \ket{\psi_n}
&= \prtq{\intmp \dd{y} \dd[n]{z} \psi_n (z_1,\dots,z_n) g(y-x) \ad(y) a(y) \ad (z_1) \dots \ad(z_n)} \ket{0},\\
&\mkern-18mu = \prtq{\intmp \dd{y} \dd[n]{z} \psi_n (z_1,\dots,z_n) g(y-x) \ad(y) \prt{\sum_i \delta (y - z_i) \prod_{j\neq i} \ad (z_j)}} \ket{0},\\
&= \sum_i \prtq{\intmp \dd[n]{z} \psi_n (z_1,\dots,z_n) g(z_i-x) \ad (z_1) \dots \ad(z_n)} \ket{0},\\
&= \prtq{\sum_i g(\hz_i - x)}\ket{\psi_n},
\end{equations}
which implies that $\sqrt{\mcN (x)}$ acts exactly as the collapse operator defined in Refs.~\cite{Dove1995Symmetric,Book_Tumulka2022Foundations}, provided we choose the proper function $g(x)$. 

In the case of a fermionic field the result is the same. It is sufficient to notice that any minus sign obtained by moving the annihilation operator $a(y)$ to the right is then countered by having to move the creation operator $\ad (y)$ to the same position. In other words, one can consider that $\ad(y) a(y) \ad(z) = \delta(y-z)\ad(y) + \ad(z) \ad(y) a(y)$.

\clearpage

\section{Derivation of the master equation\label{APPSec:MasterEquationDerivation}}

In this Appendix we show two different ways to derive Eq.~\eqref{eq:MasterEquation} of the main text.

Starting from Eq.~\eqref{APPeq:StochasticSchrodingerPoisson} and exploiting the properties of the Poisson process $\dd{N}$, we can derive the stochastic equation for the density matrix
\begin{equation}
\dd{\sigma} 
= \dyad{\dd{\psi}}{\psi} + \dyad{\psi}{\dd{\psi}} + \dyad{\dd{\psi}} 
= \int \dd[3]{x} \prtq{\dd{t} \mcQ(x) + \frac{\dd{N_x}}{\dd[3]{x}}\mcL (x)}\sigma,
\end{equation}
where
\begin{equation}
\mcQ(x) \sigma = \frac{c \gamma \mu}{2\hbar^2} \prt{2\ev{\hL^2 (x)}\sigma - \acomm{\hL^2(x)}{\sigma}},
\qquad
\mcL (x) \sigma = \frac{\hL \sigma \hL}{\ev{\hL^2}} - \sigma.
\end{equation}
Then, the master equation for $\rho_t$ can be obtained by making the ensemble average of the stochatic density matrix. We make the substitution $\dd{N_x}/\dd[3]{x} \rightarrow \dd{t} (c \gamma \mu/\hbar^2) \ev{\hL^2 (x)}$ and we get
\begin{equation}
\label{APPeq:MasterEquationNoise}
\dd{\rho_t} = \dd{t}\frac{c\mu\gamma}{\hbar^2} \int \dd[3]{x} \prtq{\hL (x) \rho_t \hL (x) - \frac{1}{2}\acomm{\hL^2 (x)}{\rho_t}}.
\end{equation}
Restoring the evolution due to the Hamiltonian $H$ of the system we get Eq.~\eqref{eq:MasterEquation} of the main text.

We can also derive Eq.~\eqref{eq:MasterEquation} directly from Eq.~\eqref{eq:TotalExactEvolution} as follows. Denoting by $\ket{\psi_t}$ the state of the system at hand at time $t$, the interaction between a collapse point located in $x_c$ at time $t$ and the system leads to
\begin{equation}
e^{-i\sqrt{\gamma} H_{x_c}/\hbar}\ket{\psi_t}\ket{0}
\simeq
\ket{\psi_t}\ket{0}
- i \frac{\sqrt{\gamma}}{\hbar} \hL (x_c)\ket{\psi_t}\ket{1} - \frac{\gamma}{2 \hbar^2}\hL^2 (x_c)\ket{\psi_t}\ket{0},
\end{equation}
where we assumed, again, that we can expand up to first order in $\gamma$. Then, the probability of getting a flash is given by
\begin{equation}
P (1) = \bra{\psi_t,0}e^{+i\sqrt{\gamma} H_{x_c}/\hbar} \dyad{1} e^{-i\sqrt{\gamma} H_{x_c}/\hbar} \ket{\psi_t,0}
\simeq \frac{\gamma}{\hbar^2} \ev{\hL^2 (x_c)}{\psi_t},
\end{equation}
and, indeed, the probability of getting no flash is just $1-P(1)$. The wavefunction evolves as follows:
\begin{equations}
\label{APPeq:CollapseGridJumps}
\ket{\psi_t} &\xrightarrow{\text{Flash}} 
\ket{\psi_t^{(1)}}\simeq 
\frac{-i \sqrt{\gamma} \hL (x_c) \ket{\psi_t}}{\hbar \sqrt{P (1)}}
\\
\ket{\psi_t} &\xrightarrow{\text{No Flash}} 
\ket{\psi_t^{(0)}}\simeq 
\frac{\prtq{1-(\gamma/2\hbar^2)\hL^2 (x_c)}\ket{\psi_t}}{\sqrt{1-P (1)}}.
\end{equations}
At the level of density matrix, this implies
\begin{multline}
\rho_t 
\rightarrow P (1) \dyad{\psi_t^{(1)}} + (1-P(1))\dyad{\psi_t^{(0)}}
\simeq
\\
\simeq \rho_t + \frac{\gamma}{\hbar^2} \prtq{\hL (x_c) \rho_t \hL (x_c) - \frac{1}{2}\acomm{\hL^2 (x_c)}{\rho_t}}.
\end{multline}

Let us assume now, as before, that the wavefunction is non-zero in a finite volume $V$. Then, in a time internal $\delta t$ we assume there are, on average, $N$ collapse events (flash or no flash) in the spacetime volume $c \delta t V$, with $N \gg 1$. Keeping again only terms up to first order in $\gamma$, we obtain (neglecting the Hamiltonian evolution of the particle)
\begin{equation}
\rho_{t + \delta t} \simeq \rho_t  + \frac{\gamma}{\hbar^2} \sum_{j=1}^N \prtq{\hL (x_j) \rho_t \hL (x_j) - \frac{1}{2}\acomm{\hL^2 (x_j)}{\rho_t}}. 
\end{equation}
Since the collapse points are uniformly distributed in space-time, it follows that we can simplify the above expression by averaging over the distribution of the collapse points in $c\delta t V$. To each collapse we associate the uniform distribution and thus we obtain
\begin{equation}
\frac{\rho_{t + \delta t} -\rho_t }{\delta t}
\simeq
\frac{c\mu\gamma}{\hbar^2} \int \dd[3]{x} \prtq{\hL (x) \rho_t \hL (x) - \frac{1}{2}\acomm{\hL^2 (x)}{\rho_t}},
\end{equation}
where $\mu = N/(c\delta t V)$ is the collapse density in spacetime. By adding the Hamiltonian evolution and making the substitution $\delta t \rightarrow \dd{t}$, we get again Eq.~\eqref{eq:MasterEquation} of the main text.

\clearpage

\section{Newtonian gravity\label{APPSec:NewtGravEnergy}}

\subsection{Energy impact of a flash on a single particle}

In this appendix, we compute the energy change of a single particle due to a flash. To do this, we choose coordinates such that the flash is situated in the origin. Then, the unitary operator $U_G$ assumes the form
\begin{equation}
U_G = \exp{i r_m \int \dd[3]{y}\frac{f_G(\abs{y})}{\abs{\hx-y}}},
\qquad
r_m \equiv \frac{G m_R m}{\lambda_{\rm GRW} \hbar},
\end{equation}
where we exploited the fact that, for a single particle, $\hM(x) = m \delta (\hx - x)$. In spherical coordinates, this unitary operator changes a wavefunction $\psi(r,\theta,\phi)$ as follows
\begin{multline}
U_G \rightarrow \exp \prtgB{i r_m \int_0^{\infty} \dd{r_y} \int_0^{\pi} \dd{\theta_y} \int_{0}^{2\pi} \dd{\phi_y} 
\times \\ \times 
\frac{r_y^2 \sin(\theta_y) f_G (r_y)}{\sqrt{r^2 + r_y^2 -2 r r_y \prtq{\sin(\theta)\sin(\theta_y)\cos(\phi - \phi_y)+\cos(\theta)\cos(\theta_y)}}}}.
\end{multline}
However, despite the appearances, the above function is only a function of $r$ and does not depend on $\theta$ or $\phi$. So, we can write the unitary operator in spherical coordinate representation as $U_G (r) = \exp[i r_m F(r)]$. Setting $\theta=0$, $F(r)$ assumes the following form (along with its derivative)
\begin{equations}
F(r) 
&= 2 \pi \int_0^{\infty} \dd{r_y} \frac{r_y f_G (r_y)}{r}\prt{r+r_y -\abs{r-r_y}},\\
&= 4 \pi \prtq{\frac{1}{r}\int_0^r \dd{r_y} r_y^2 f_G (r_y) + \int_r^{\infty} \dd{r_y}  r_y f_G (r_y)},\\
F'(r) &= - \frac{4 \pi}{r^2} \int_0^r \dd{r_y} r_y^2 f_G (r_y).
%F'' (r) &= -4\pi f_G (r) + \frac{8 \pi}{r^3} \int_0^r \dd{r_y} r_y^2 f_G (r_y).
\end{equations}

The kinetic energy after a gravitational pull can then be calculated as follows
\begin{equation}
\ev{\frac{\hp^2}{2m}} = -\frac{\hbar^2}{2m}
\int_0^{\infty} \dd{r} \int_0^{\pi} \dd{\theta} \int_{0}^{2\pi} \dd{\phi} r^2 \sin(\theta) \prtq{\psi^* (r,\theta,\phi) U_G^*(r) \nabla^2 U_G (r)\psi(r,\theta,\phi)}.
\end{equation}
One can then see that
\begin{equations}
\nabla^2 U_G \psi 
&= \frac{2 \psi}{r}\partial_r U_G + \psi \partial^2_r U_G + 2 \prt{\partial_r \psi}\prt{\partial_r U_G} + U_G \nabla^2 \psi,\\
&= U_G \prtq{- r_m^2 \prt{F'(r)}^2 \psi + i r_m \prt{F''(r) \psi + \frac{2}{r} F'(r) \psi + 2 F'(r) \partial_r \psi}} + U_G \nabla^2 \psi,\\
&= U_G \prtq{- r_m^2 \prt{F'(r)}^2 \psi + i r_m \frac{1}{r^2}\partial_r \prt{r^2 F'(r) \psi} + i r_m F'(r) \partial_r \psi } + U_G \nabla^2 \psi,\\
\end{equations}
which leads to
\begin{multline}
\ev{\frac{\hp^2}{2m}} = -\frac{\hbar^2}{2m}
\int_0^{\infty} \dd{r} \int_0^{\pi} \dd{\theta} \int_{0}^{2\pi} \dd{\phi} \sin(\theta) \times \\ \times
\prtq{r^2 \psi^*\nabla^2\psi - r_m^2 r^2 \prt{F'(r)}^2 \abs{\psi}^2 + i r_m \psi^* \partial_r \prt{r^2 F'(r) \psi} + i r_m r^2 F'(r) \psi^*\partial_r \psi }.
\end{multline}
The first term in the integral is just the energy prior to the gravitational pull while the other terms contain the energy due to it. We can integrate by parts the third term and see that it is minus the complex conjugate of the fourth. Therefore, we get
\begin{multline}
\ev{\frac{\hp^2}{2m}} = -\frac{\hbar^2}{2m}
\int_0^{\infty} \dd{r} \int_0^{\pi} \dd{\theta} \int_{0}^{2\pi} \dd{\phi} \prtgB{ \\
r^2 \sin(\theta)
\prtq{\psi^*\nabla^2\psi - r_m^2 \prt{F'(r)}^2 \abs{\psi}^2 - 2 r_m F'(r) \Im{\psi^* \partial_r \psi}}}.
\end{multline}
One immediately see that if $F(r)=1/r$, the second term in the above integral may lead to a divergence. However, with the smearing of the collapse point as the gravitational source, this does not happen. For example, one could take $f_G (r) = (\pi r_G)^{-3/2} \exp[-r^2/r_G^2]$ to avoid this kind of divergences.

\subsection{Single particle decoherence dynamics}

Let us analyse the dynamics of a single particle of mass $m$ governed by Eq.~\eqref{eq:GravitationalMasterEquation}. Since we are dealing with a single particle, the mass operator $\hM (x)$ becomes $\hM (x) \rightarrow m \delta(x-\hx)$ so that the unitary $U_G (x)$ and the collapse operator $\hL (x)$ now read
\begin{equation}
U_G (s) = \exp[i r_m\int \dd[3]z \frac{f_G(\abs{z-s})}{{\abs{\hx-z}}}],
\quad
r_m   \equiv \frac{G m_R m}{\lambda_{\rm GRW}\hbar},
\qquad
\hL(s) = \sqrt{\frac{m}{m_R}}f_C (\abs{\hx-s}),
\end{equation}
where $r_m  $ is a new length scale associated to the gravitational effect of a flash and associated to the mass $m$ of the particle, $f_C (\abs{x})$ is a real function, usually taken to be a Gaussian, characterised by the collapse radius $r_C$~\cite{Bassi2013Models,Tilloy2018GRWGravity}, while $f_G (\abs{x})$ is also a real function but characterised by a radius $r_G$. $r_G$ would be a new parameter but one could choose to set it equal to $r_C$. We leave it as a separate parameter for the sake of generality. For standard quantum particles it holds that $r_m \ll r_C$~\cite{Tilloy2018GRWGravity}.

In coordinate representation, neglecting the standard Hamiltonian of the particle, we get the single particle master equation
\begin{equation}
\label{APPeq:MasterEquationSingleParticle}
\dv{t} \mel{x}{\rho_t}{y} = \lambda_{\rm GRW} \frac{m}{m_R}\Gamma(x,y)\mel{x}{\rho_t}{y},
\end{equation}
where
\begin{multline}
\Gamma(x,y) = \int \dd[3]{s} \prtgB{
- \frac{f_C^2(\abs{x-s})}{2}- \frac{f_C^2(\abs{y-s})}{2}
+\\+
\exp[i r_m   \int \dd[3]{z} \prt{\frac{f_G(\abs{z-s})}{\abs{x-z}}-\frac{f_G(\abs{z-s})}{\abs{y-z}}}]f_C(\abs{x-s})f_C(\abs{y-s})}.
\end{multline}
Notice how, by taking the function $f_C (x)$ to be the standard GRW choice, i.e., $f_C (x) = (\pi r_C^2)^{-3/4}\exp[-x^2/(2 r_C^2)]$, the equation for $\Gamma(x,y)-1$ becomes very similar to the one obtained in Ref.~\cite{Tilloy2018GRWGravity}, Eqs.~(10) and~(11)\footnote{We recall that $\int \dd[3]{x} f_C^2 (x) = 1$ with the GRW choice.}.

Independently of the choice of $f_C (\abs{x})$, $\Gamma(x,x)=0$ and $\Gamma(x,y)=\Gamma^*(y,x)$. We now prove that $\Gamma(x,y) \in \mathbb{R}$ and that $\Gamma(x,y)$ is actually a function of the single variable $d=\abs{x-y}/2$. First, we observe that the last two terms in the integral are real so we only need to prove that the first term gives rise to a real term. Secondly, we define $F(\abs{x-s}) = \int \dd[3]{z} f_G(\abs{z-s})/\abs{x-z}$, where we can see that the result of the integral only depends on $\abs{x-s}$ because we can choose, for each $x$ and $s$, a coordinate system where $x=0$ and $s$ has just one nonzero coordinate equal to $\abs{x-s}$. After noticing this, we make the change of variable $s'= s - (x+y)/2$ which is a translation of the origin. Then, we rotate the coordinate system so that for a given couple of points $(x,y)$ the point $d = (x-y)/2$ is along the $z$-axis with its $z$-coordinate being positive (or zero). It follows that, in cylindrical coordinates and defining $l(z,r,d) \equiv \sqrt{(z+d)^2+r^2}$, the above integral reads
\begin{multline}
\Gamma(x,y) = 2 \pi \int_0^{\infty} \dd{r} r \prtgB{ - \intmp \dd{z}\frac{f_C^2\prt{l(z,r,-d)}}{2} - \intmp \dd{z}\frac{f_C^2\prt{l(z,r,d)}}{2} +\\
+\int_{0}^{+\infty} \dd{z}\exp{i r_m  \prtq{F \prt{l(z,r,d)}-F \prt{l(z,r,-d)}}}f_C\prt{l(z,r,d)}f_C\prt{l(z,r,-d)} +\\
+\int_{0}^{+\infty} \dd{z}\exp{i r_m  \prtq{F \prt{l(z,r,-d)}-F \prt{l(z,r,d)}}}f_C\prt{l(z,r,-d)}f_C\prt{l(z,r,d)}
},
\end{multline}
where we first splitted in two the integral in the $z$ variable and then, in the second line, we made the substitution $z \rightarrow -z$ and re-ordered the integration extremes. We see that second and third line are the complex conjugate of each other. It follows then that $\Gamma(x,y)$ is real. Finally, with a slight abuse of notation, we can write
\begin{multline}
\label{APPeq:GammaFunction}
\Gamma(d) = 2 \pi \int_{0}^{\infty} \dd{r} r \intmp \dd{z} \prtgB{-\frac{f_C^2\prt{l(z,r,-d)}}{2}-\frac{f_C^2\prt{l(z,r,-d)}}{2} +
\\
\cos \prt{r_m  \prtq{F \prt{l(z,r,-d)}-F \prt{l(z,r,-d)}}}f_C\prt{l(z,r,-d)}f_C\prt{l(z,r,-d)}}.
\end{multline}

Thus, as in Ref.~\cite{Tilloy2018GRWGravity}, the only effect of self gravity, when neglecting the particle bare Hamiltonian, is to increase the rate of dephasing in position basis. Moreover, the dephasing rate of the density matrix $\rho (x,y)$ only depends on the distance $d = \abs{x-y}/2$. 

\subsection{Single particle short distance decoherence behaviour}

We now turn to study the modifications to the behaviour at short distances due to the presence of the gravitational term. This calculation purpose is just to show the emergence of a dependence of the decoherence on the inverse of a power of $\lambda_{\rm GRW}$, thus hinting at the possibility that a full model of such Newtonian gravity can pose testable, in principle, experimental bound from below for $\lambda_{\rm GRW}$. To make this proof of principle calculation, we follow again Ref.~\cite{Tilloy2018GRWGravity}, i.e., we choose $F(\abs{x-s}) = 1/\abs{x-s}$ and we use the GRW collapse operator $f_C (x) = (\pi r_C^2)^{-3/4}\exp[-x^2/(2 r_C^2)]$. We, as expected, obtain the same qualitative result.

The first step is to write Eq.~\eqref{APPeq:GammaFunction} in spherical coordinates:
\begin{multline}
\label{APPeq:GammaFunctionSphericalGRW}
\Gamma (d) + 1 = 2 \pi e^{-d^2/r_C^2}\int_{0}^{\infty} \dd{r} r^2 \int_0^{\pi} \dd{\theta} 
\times \\ \times
\sin(\theta) e^{-r^2/r_C^2}\cos \prt{\frac{r_m  }{\sqrt{r^2 + 2 d r \cos(\theta) + d^2}}-\frac{r_m  }{\sqrt{r^2 - 2 d r \cos(\theta) + d^2}}}.
\end{multline}
The functions inside the integral of Eq.~\eqref{APPeq:GammaFunctionSphericalGRW} are all bounded so that we can choose a small length $\epsilon$ such that the integration for $r \in [0,\epsilon]$ gives an almost vanishing contribution. We then expand, at first order, the argument of the cosine for $d \ll \epsilon$ and get
\begin{equation}
\label{APPeq:CosineArgumentExpansion}
\cos \prt{\frac{r_m  }{\sqrt{r^2 + 2 d r \cos(\theta) + d^2}}-\frac{r_m  }{\sqrt{r^2 - 2 d r \cos(\theta) + d^2}}}
\simeq
\cos(\frac{2 r_m   \cos(\theta) d}{r^2}),
\end{equation}
which can be integrated as follows
\begin{equation}
\int_0^\pi \dd{\theta} \sin(\theta )\cos(\frac{2 r_m   \cos(\theta) d}{r^2})
=
-\prtq{\frac{r^2}{2 r_m  d} \sin(\frac{2 r_m   d \cos(\theta)}{r^2})}^\pi_0
=
\frac{r^2}{r_m  d} \sin(\frac{2 r_m   d}{r^2}).
\end{equation}
Substituting this back leads to an integral that can be solved analytically (with $\epsilon =0$) but whose solution is quite cumbersome. Expanding this solution to second order in $d$ gives
\begin{equation}
\Gamma(d) \simeq - \frac{32}{15}\frac{r_m^{3/2}  }{r_C^3}d^{3/2} - \frac{1}{r_C^2}\prt{1-\frac{8}{3}\frac{r_m^2  }{r_C^2}}d^2.
\end{equation}
The first leading order is much more important than the second when $d \ll r_m^3  /r_C^2, r_C^2/r_m  $. The first leading order agrees with the result of Ref.~\cite{Tilloy2018GRWGravity}\footnote{Eq.~(12) of Ref.~\cite{Tilloy2018GRWGravity} gives a linear scaling. Private communication with the author confirms it should have been $d^{3/2}$.}.

\clearpage

\end{appendices}


\begin{thebibliography}{49}%
\makeatletter
\providecommand \@ifxundefined [1]{%
 \@ifx{#1\undefined}
}%
\providecommand \@ifnum [1]{%
 \ifnum #1\expandafter \@firstoftwo
 \else \expandafter \@secondoftwo
 \fi
}%
\providecommand \@ifx [1]{%
 \ifx #1\expandafter \@firstoftwo
 \else \expandafter \@secondoftwo
 \fi
}%
\providecommand \natexlab [1]{#1}%
\providecommand \enquote  [1]{``#1''}%
\providecommand \bibnamefont  [1]{#1}%
\providecommand \bibfnamefont [1]{#1}%
\providecommand \citenamefont [1]{#1}%
\providecommand \href@noop [0]{\@secondoftwo}%
\providecommand \href [0]{\begingroup \@sanitize@url \@href}%
\providecommand \@href[1]{\@@startlink{#1}\@@href}%
\providecommand \@@href[1]{\endgroup#1\@@endlink}%
\providecommand \@sanitize@url [0]{\catcode `\\12\catcode `\$12\catcode
  `\&12\catcode `\#12\catcode `\^12\catcode `\_12\catcode `\%12\relax}%
\providecommand \@@startlink[1]{}%
\providecommand \@@endlink[0]{}%
\providecommand \url  [0]{\begingroup\@sanitize@url \@url }%
\providecommand \@url [1]{\endgroup\@href {#1}{\urlprefix }}%
\providecommand \urlprefix  [0]{URL }%
\providecommand \Eprint [0]{\href }%
\providecommand \doibase [0]{https://doi.org/}%
\providecommand \selectlanguage [0]{\@gobble}%
\providecommand \bibinfo  [0]{\@secondoftwo}%
\providecommand \bibfield  [0]{\@secondoftwo}%
\providecommand \translation [1]{[#1]}%
\providecommand \BibitemOpen [0]{}%
\providecommand \bibitemStop [0]{}%
\providecommand \bibitemNoStop [0]{.\EOS\space}%
\providecommand \EOS [0]{\spacefactor3000\relax}%
\providecommand \BibitemShut  [1]{\csname bibitem#1\endcsname}%
\let\auto@bib@innerbib\@empty
%</preamble>
\bibitem [{\citenamefont {Bassi}\ and\ \citenamefont
  {Ghirardi}(2003)}]{Bassi2003Dynamical}%
  \BibitemOpen
  \bibfield  {author} {\bibinfo {author} {\bibfnamefont {A.}~\bibnamefont
  {Bassi}}\ and\ \bibinfo {author} {\bibfnamefont {G.}~\bibnamefont
  {Ghirardi}},\ }\bibfield  {title} {\bibinfo {title} {Dynamical reduction
  models},\ }\href
  {https://doi.org/https://doi.org/10.1016/S0370-1573(03)00103-0} {\bibfield
  {journal} {\bibinfo  {journal} {Physics Reports}\ }\textbf {\bibinfo {volume}
  {379}},\ \bibinfo {pages} {257} (\bibinfo {year} {2003})}\BibitemShut
  {NoStop}%
\bibitem [{\citenamefont {Bassi}\ \emph {et~al.}(2013)\citenamefont {Bassi},
  \citenamefont {Lochan}, \citenamefont {Satin}, \citenamefont {Singh},\ and\
  \citenamefont {Ulbricht}}]{Bassi2013Models}%
  \BibitemOpen
  \bibfield  {author} {\bibinfo {author} {\bibfnamefont {A.}~\bibnamefont
  {Bassi}}, \bibinfo {author} {\bibfnamefont {K.}~\bibnamefont {Lochan}},
  \bibinfo {author} {\bibfnamefont {S.}~\bibnamefont {Satin}}, \bibinfo
  {author} {\bibfnamefont {T.~P.}\ \bibnamefont {Singh}},\ and\ \bibinfo
  {author} {\bibfnamefont {H.}~\bibnamefont {Ulbricht}},\ }\bibfield  {title}
  {\bibinfo {title} {Models of wave-function collapse, underlying theories, and
  experimental tests},\ }\href {https://doi.org/10.1103/RevModPhys.85.471}
  {\bibfield  {journal} {\bibinfo  {journal} {Rev. Mod. Phys.}\ }\textbf
  {\bibinfo {volume} {85}},\ \bibinfo {pages} {471} (\bibinfo {year}
  {2013})}\BibitemShut {NoStop}%
\bibitem [{\citenamefont {Bassi}\ \emph {et~al.}(2023)\citenamefont {Bassi},
  \citenamefont {Dorato},\ and\ \citenamefont
  {Ulbricht}}]{Bassi2023CollapseModels}%
  \BibitemOpen
  \bibfield  {author} {\bibinfo {author} {\bibfnamefont {A.}~\bibnamefont
  {Bassi}}, \bibinfo {author} {\bibfnamefont {M.}~\bibnamefont {Dorato}},\ and\
  \bibinfo {author} {\bibfnamefont {H.}~\bibnamefont {Ulbricht}},\ }\bibfield
  {title} {\bibinfo {title} {Collapse models: A theoretical, experimental and
  philosophical review},\ }\href {https://doi.org/10.3390/e25040645} {\bibfield
   {journal} {\bibinfo  {journal} {Entropy}\ }\textbf {\bibinfo {volume}
  {25}},\ \bibinfo {pages} {645} (\bibinfo {year} {2023})}\BibitemShut
  {NoStop}%
\bibitem [{\citenamefont {Josset}\ \emph {et~al.}(2017)\citenamefont {Josset},
  \citenamefont {Perez},\ and\ \citenamefont
  {Sudarsky}}]{Josset2017DarkEnergy}%
  \BibitemOpen
  \bibfield  {author} {\bibinfo {author} {\bibfnamefont {T.}~\bibnamefont
  {Josset}}, \bibinfo {author} {\bibfnamefont {A.}~\bibnamefont {Perez}},\ and\
  \bibinfo {author} {\bibfnamefont {D.}~\bibnamefont {Sudarsky}},\ }\bibfield
  {title} {\bibinfo {title} {Dark energy from violation of energy
  conservation},\ }\href {https://doi.org/10.1103/PhysRevLett.118.021102}
  {\bibfield  {journal} {\bibinfo  {journal} {Phys. Rev. Lett.}\ }\textbf
  {\bibinfo {volume} {118}},\ \bibinfo {pages} {021102} (\bibinfo {year}
  {2017})}\BibitemShut {NoStop}%
\bibitem [{\citenamefont {Karolyhazy}(1966)}]{Karolyhazy1966Gravitation}%
  \BibitemOpen
  \bibfield  {author} {\bibinfo {author} {\bibfnamefont {F.}~\bibnamefont
  {Karolyhazy}},\ }\bibfield  {title} {\bibinfo {title} {Gravitation and
  quantum mechanics of macroscopic objects},\ }\href
  {https://doi.org/10.1007/BF02717926} {\bibfield  {journal} {\bibinfo
  {journal} {Il Nuovo Cimento A (1965-1970)}\ }\textbf {\bibinfo {volume}
  {42}},\ \bibinfo {pages} {390} (\bibinfo {year} {1966})}\BibitemShut
  {NoStop}%
\bibitem [{\citenamefont {K\'{a}rolyh\'{a}zy}\ \emph
  {et~al.}(1986)\citenamefont {K\'{a}rolyh\'{a}zy}, \citenamefont {Frenkel},\
  and\ \citenamefont {Luk\'{a}cs}}]{Karolyhazy1986possible}%
  \BibitemOpen
  \bibfield  {author} {\bibinfo {author} {\bibfnamefont {F.}~\bibnamefont
  {K\'{a}rolyh\'{a}zy}}, \bibinfo {author} {\bibfnamefont {A.}~\bibnamefont
  {Frenkel}},\ and\ \bibinfo {author} {\bibfnamefont {B.}~\bibnamefont
  {Luk\'{a}cs}},\ }\bibfield  {title} {\bibinfo {title} {On the possible role
  of gravity in the reduction of the wave function},\ }in\ \href
  {https://philpapers.org/rec/KROOTP} {\emph {\bibinfo {booktitle} {Quantum
  Concepts in Space and Time}}},\ \bibinfo {editor} {edited by\ \bibinfo
  {editor} {\bibfnamefont {R.}~\bibnamefont {Penrose}}\ and\ \bibinfo {editor}
  {\bibfnamefont {C.~J.}\ \bibnamefont {Isham}}}\ (\bibinfo  {publisher}
  {Oxford University Press},\ \bibinfo {address} {New York},\ \bibinfo {year}
  {1986})\ pp.\ \bibinfo {pages} {1--109}\BibitemShut {NoStop}%
\bibitem [{\citenamefont {Diósi}(1987)}]{Diosi1987Universal}%
  \BibitemOpen
  \bibfield  {author} {\bibinfo {author} {\bibfnamefont {L.}~\bibnamefont
  {Diósi}},\ }\bibfield  {title} {\bibinfo {title} {A universal master
  equation for the gravitational violation of quantum mechanics},\ }\href
  {https://doi.org/https://doi.org/10.1016/0375-9601(87)90681-5} {\bibfield
  {journal} {\bibinfo  {journal} {Physics Letters A}\ }\textbf {\bibinfo
  {volume} {120}},\ \bibinfo {pages} {377} (\bibinfo {year}
  {1987})}\BibitemShut {NoStop}%
\bibitem [{\citenamefont {Di\'osi}(1989)}]{Diosi1989Models}%
  \BibitemOpen
  \bibfield  {author} {\bibinfo {author} {\bibfnamefont {L.}~\bibnamefont
  {Di\'osi}},\ }\bibfield  {title} {\bibinfo {title} {Models for universal
  reduction of macroscopic quantum fluctuations},\ }\href
  {https://doi.org/10.1103/PhysRevA.40.1165} {\bibfield  {journal} {\bibinfo
  {journal} {Phys. Rev. A}\ }\textbf {\bibinfo {volume} {40}},\ \bibinfo
  {pages} {1165} (\bibinfo {year} {1989})}\BibitemShut {NoStop}%
\bibitem [{\citenamefont {Penrose}(1996)}]{Penrose1996gravity}%
  \BibitemOpen
  \bibfield  {author} {\bibinfo {author} {\bibfnamefont {R.}~\bibnamefont
  {Penrose}},\ }\bibfield  {title} {\bibinfo {title} {On gravity's role in
  quantum state reduction},\ }\href
  {https://link.springer.com/article/10.1007/bf02105068} {\bibfield  {journal}
  {\bibinfo  {journal} {General relativity and gravitation}\ }\textbf {\bibinfo
  {volume} {28}},\ \bibinfo {pages} {581} (\bibinfo {year} {1996})}\BibitemShut
  {NoStop}%
\bibitem [{\citenamefont {Bassi}\ \emph {et~al.}(2017)\citenamefont {Bassi},
  \citenamefont {Großardt},\ and\ \citenamefont
  {Ulbricht}}]{Bassi2017Gravitational}%
  \BibitemOpen
  \bibfield  {author} {\bibinfo {author} {\bibfnamefont {A.}~\bibnamefont
  {Bassi}}, \bibinfo {author} {\bibfnamefont {A.}~\bibnamefont {Großardt}},\
  and\ \bibinfo {author} {\bibfnamefont {H.}~\bibnamefont {Ulbricht}},\
  }\bibfield  {title} {\bibinfo {title} {Gravitational decoherence},\ }\href
  {https://doi.org/10.1088/1361-6382/aa864f} {\bibfield  {journal} {\bibinfo
  {journal} {Classical and Quantum Gravity}\ }\textbf {\bibinfo {volume}
  {34}},\ \bibinfo {pages} {193002} (\bibinfo {year} {2017})}\BibitemShut
  {NoStop}%
\bibitem [{\citenamefont {Gasbarri}\ \emph {et~al.}(2017)\citenamefont
  {Gasbarri}, \citenamefont {Toro\ifmmode~\check{s}\else \v{s}\fi{}},
  \citenamefont {Donadi},\ and\ \citenamefont {Bassi}}]{Gasbarri2017Gravity}%
  \BibitemOpen
  \bibfield  {author} {\bibinfo {author} {\bibfnamefont {G.}~\bibnamefont
  {Gasbarri}}, \bibinfo {author} {\bibfnamefont {M.}~\bibnamefont
  {Toro\ifmmode~\check{s}\else \v{s}\fi{}}}, \bibinfo {author} {\bibfnamefont
  {S.}~\bibnamefont {Donadi}},\ and\ \bibinfo {author} {\bibfnamefont
  {A.}~\bibnamefont {Bassi}},\ }\bibfield  {title} {\bibinfo {title} {Gravity
  induced wave function collapse},\ }\href
  {https://doi.org/10.1103/PhysRevD.96.104013} {\bibfield  {journal} {\bibinfo
  {journal} {Phys. Rev. D}\ }\textbf {\bibinfo {volume} {96}},\ \bibinfo
  {pages} {104013} (\bibinfo {year} {2017})}\BibitemShut {NoStop}%
\bibitem [{\citenamefont {Tilloy}\ and\ \citenamefont
  {Di\'osi}(2016)}]{Tilloy2016CSLGravity}%
  \BibitemOpen
  \bibfield  {author} {\bibinfo {author} {\bibfnamefont {A.}~\bibnamefont
  {Tilloy}}\ and\ \bibinfo {author} {\bibfnamefont {L.}~\bibnamefont
  {Di\'osi}},\ }\bibfield  {title} {\bibinfo {title} {Sourcing semiclassical
  gravity from spontaneously localized quantum matter},\ }\href
  {https://doi.org/10.1103/PhysRevD.93.024026} {\bibfield  {journal} {\bibinfo
  {journal} {Phys. Rev. D}\ }\textbf {\bibinfo {volume} {93}},\ \bibinfo
  {pages} {024026} (\bibinfo {year} {2016})}\BibitemShut {NoStop}%
\bibitem [{\citenamefont {Tilloy}(2018)}]{Tilloy2018GRWGravity}%
  \BibitemOpen
  \bibfield  {author} {\bibinfo {author} {\bibfnamefont {A.}~\bibnamefont
  {Tilloy}},\ }\bibfield  {title} {\bibinfo {title} {Ghirardi-rimini-weber
  model with massive flashes},\ }\href
  {https://doi.org/10.1103/PhysRevD.97.021502} {\bibfield  {journal} {\bibinfo
  {journal} {Phys. Rev. D}\ }\textbf {\bibinfo {volume} {97}},\ \bibinfo
  {pages} {021502} (\bibinfo {year} {2018})}\BibitemShut {NoStop}%
\bibitem [{\citenamefont {Carlesso}\ and\ \citenamefont
  {Donadi}(2019)}]{Carlesso2019Collapse}%
  \BibitemOpen
  \bibfield  {author} {\bibinfo {author} {\bibfnamefont {M.}~\bibnamefont
  {Carlesso}}\ and\ \bibinfo {author} {\bibfnamefont {S.}~\bibnamefont
  {Donadi}},\ }\bibfield  {title} {\bibinfo {title} {Collapse models: Main
  properties and the state of art of the experimental tests},\ }in\ \href
  {https://doi.org/https://doi.org/10.1007/978-3-030-31146-9_1} {\emph
  {\bibinfo {booktitle} {Advances in Open Systems and Fundamental Tests of
  Quantum Mechanics}}},\ \bibinfo {editor} {edited by\ \bibinfo {editor}
  {\bibfnamefont {B.}~\bibnamefont {Vacchini}}, \bibinfo {editor}
  {\bibfnamefont {H.-P.}\ \bibnamefont {Breuer}},\ and\ \bibinfo {editor}
  {\bibfnamefont {A.}~\bibnamefont {Bassi}}}\ (\bibinfo  {publisher} {Springer
  International Publishing},\ \bibinfo {address} {Cham},\ \bibinfo {year}
  {2019})\ pp.\ \bibinfo {pages} {1--13}\BibitemShut {NoStop}%
\bibitem [{\citenamefont {Carlesso}\ \emph {et~al.}(2022)\citenamefont
  {Carlesso}, \citenamefont {Donadi}, \citenamefont {Ferialdi}, \citenamefont
  {Paternostro}, \citenamefont {Ulbricht},\ and\ \citenamefont
  {Bassi}}]{Carlesso2022Present}%
  \BibitemOpen
  \bibfield  {author} {\bibinfo {author} {\bibfnamefont {M.}~\bibnamefont
  {Carlesso}}, \bibinfo {author} {\bibfnamefont {S.}~\bibnamefont {Donadi}},
  \bibinfo {author} {\bibfnamefont {L.}~\bibnamefont {Ferialdi}}, \bibinfo
  {author} {\bibfnamefont {M.}~\bibnamefont {Paternostro}}, \bibinfo {author}
  {\bibfnamefont {H.}~\bibnamefont {Ulbricht}},\ and\ \bibinfo {author}
  {\bibfnamefont {A.}~\bibnamefont {Bassi}},\ }\bibfield  {title} {\bibinfo
  {title} {Present status and future challenges of non-interferometric tests of
  collapse models},\ }\href
  {https://www.nature.com/articles/s41567-021-01489-5#article-info} {\bibfield
  {journal} {\bibinfo  {journal} {Nature Physics}\ }\textbf {\bibinfo {volume}
  {18}},\ \bibinfo {pages} {243} (\bibinfo {year} {2022})}\BibitemShut
  {NoStop}%
\bibitem [{\citenamefont {Snoke}(2021)}]{Snoke2021Collapse}%
  \BibitemOpen
  \bibfield  {author} {\bibinfo {author} {\bibfnamefont {D.}~\bibnamefont
  {Snoke}},\ }\bibfield  {title} {\bibinfo {title} {A model of spontaneous
  collapse with energy conservation},\ }\href
  {https://doi.org/10.1007/s10701-021-00507-z} {\bibfield  {journal} {\bibinfo
  {journal} {Foundations of Physics}\ }\textbf {\bibinfo {volume} {51}},\
  \bibinfo {pages} {100} (\bibinfo {year} {2021})}\BibitemShut {NoStop}%
\bibitem [{\citenamefont {Dove}\ and\ \citenamefont
  {Squires}(1995)}]{Dove1995Symmetric}%
  \BibitemOpen
  \bibfield  {author} {\bibinfo {author} {\bibfnamefont {C.}~\bibnamefont
  {Dove}}\ and\ \bibinfo {author} {\bibfnamefont {E.~J.}\ \bibnamefont
  {Squires}},\ }\bibfield  {title} {\bibinfo {title} {Symmetric versions of
  explicit wavefunction collapse models},\ }\href
  {https://cds.cern.ch/record/272380/files/SCAN-9411346.pdf} {\bibfield
  {journal} {\bibinfo  {journal} {Foundations of Physics}\ }\textbf {\bibinfo
  {volume} {25}},\ \bibinfo {pages} {1267} (\bibinfo {year}
  {1995})}\BibitemShut {NoStop}%
\bibitem [{\citenamefont
  {Tumulka}(2006{\natexlab{a}})}]{Tumulka2006spontaneous}%
  \BibitemOpen
  \bibfield  {author} {\bibinfo {author} {\bibfnamefont {R.}~\bibnamefont
  {Tumulka}},\ }\bibfield  {title} {\bibinfo {title} {On spontaneous wave
  function collapse and quantum field theory},\ }\href
  {https://doi.org/10.1098/rspa.2005.1636} {\bibfield  {journal} {\bibinfo
  {journal} {Proceedings of the Royal Society A: Mathematical, Physical and
  Engineering Sciences}\ }\textbf {\bibinfo {volume} {462}},\ \bibinfo {pages}
  {1897} (\bibinfo {year} {2006}{\natexlab{a}})}\BibitemShut {NoStop}%
\bibitem [{\citenamefont {Bell}(2001)}]{Bell1987JumpsV1}%
  \BibitemOpen
  \bibfield  {author} {\bibinfo {author} {\bibfnamefont {J.~S.}\ \bibnamefont
  {Bell}},\ }\bibfield  {title} {\bibinfo {title} {Are there quantum jumps?},\
  }in\ \href {https://doi.org/10.1142/4757} {\emph {\bibinfo {booktitle} {John
  S Bell on the Foundations of Quantum Mechanics}}}\ (\bibinfo  {publisher}
  {WORLD SCIENTIFIC},\ \bibinfo {year} {2001})\BibitemShut {NoStop}%
\bibitem [{\citenamefont {Bell}(2004)}]{Bell1987JumpsV2}%
  \BibitemOpen
  \bibfield  {author} {\bibinfo {author} {\bibfnamefont {J.~S.}\ \bibnamefont
  {Bell}},\ }\bibfield  {title} {\bibinfo {title} {Are there quantum jumps?},\
  }in\ \href {https://doi.org/10.1017/CBO9780511815676} {\emph {\bibinfo
  {booktitle} {Speakable and unspeakable in quantum mechanics: Collected papers
  on quantum philosophy}}}\ (\bibinfo  {publisher} {Cambridge university
  press},\ \bibinfo {year} {2004})\BibitemShut {NoStop}%
\bibitem [{\citenamefont
  {Tumulka}(2006{\natexlab{b}})}]{Tumulka2006relativistic}%
  \BibitemOpen
  \bibfield  {author} {\bibinfo {author} {\bibfnamefont {R.}~\bibnamefont
  {Tumulka}},\ }\bibfield  {title} {\bibinfo {title} {A relativistic version of
  the ghirardi--rimini--weber model},\ }\href
  {https://link.springer.com/article/10.1007/s10955-006-9227-3} {\bibfield
  {journal} {\bibinfo  {journal} {Journal of Statistical Physics}\ }\textbf
  {\bibinfo {volume} {125}},\ \bibinfo {pages} {821} (\bibinfo {year}
  {2006}{\natexlab{b}})}\BibitemShut {NoStop}%
\bibitem [{\citenamefont {Esfeld}\ and\ \citenamefont
  {Gisin}(2014)}]{Esfeld2014GRW}%
  \BibitemOpen
  \bibfield  {author} {\bibinfo {author} {\bibfnamefont {M.}~\bibnamefont
  {Esfeld}}\ and\ \bibinfo {author} {\bibfnamefont {N.}~\bibnamefont {Gisin}},\
  }\bibfield  {title} {\bibinfo {title} {The grw flash theory: A relativistic
  quantum ontology of matter in space-time?},\ }\href
  {https://doi.org/10.1086/675730} {\bibfield  {journal} {\bibinfo  {journal}
  {Philosophy of Science}\ }\textbf {\bibinfo {volume} {81}},\ \bibinfo {pages}
  {248–264} (\bibinfo {year} {2014})}\BibitemShut {NoStop}%
\bibitem [{\citenamefont {Tumulka}(2022)}]{Book_Tumulka2022Foundations}%
  \BibitemOpen
  \bibfield  {author} {\bibinfo {author} {\bibfnamefont {R.}~\bibnamefont
  {Tumulka}},\ }\href {https://books.google.com.sg/books?id=VR4tzwEACAAJ}
  {\emph {\bibinfo {title} {Foundations of Quantum Mechanics}}},\ Lecture Notes
  in Physics\ (\bibinfo  {publisher} {Springer International Publishing},\
  \bibinfo {year} {2022})\BibitemShut {NoStop}%
\bibitem [{\citenamefont {Wiseman}\ and\ \citenamefont
  {Milburn}(2009)}]{Book_Wiseman2009Measurement}%
  \BibitemOpen
  \bibfield  {author} {\bibinfo {author} {\bibfnamefont {H.~M.}\ \bibnamefont
  {Wiseman}}\ and\ \bibinfo {author} {\bibfnamefont {G.~J.}\ \bibnamefont
  {Milburn}},\ }\href {https://doi.org/10.1017/CBO9780511813948} {\emph
  {\bibinfo {title} {Quantum Measurement and Control}}}\ (\bibinfo  {publisher}
  {Cambridge University Press},\ \bibinfo {year} {2009})\BibitemShut {NoStop}%
\bibitem [{\citenamefont {Penrose}(2014)}]{Penrose2014Gravitization}%
  \BibitemOpen
  \bibfield  {author} {\bibinfo {author} {\bibfnamefont {R.}~\bibnamefont
  {Penrose}},\ }\bibfield  {title} {\bibinfo {title} {On the gravitization of
  quantum mechanics 1: Quantum state reduction},\ }\href
  {https://doi.org/10.1007/s10701-013-9770-0} {\bibfield  {journal} {\bibinfo
  {journal} {Foundations of Physics}\ }\textbf {\bibinfo {volume} {44}},\
  \bibinfo {pages} {557} (\bibinfo {year} {2014})}\BibitemShut {NoStop}%
\bibitem [{\citenamefont {Oppenheim}(2021)}]{Oppenheim2021postquantum}%
  \BibitemOpen
  \bibfield  {author} {\bibinfo {author} {\bibfnamefont {J.}~\bibnamefont
  {Oppenheim}},\ }\bibfield  {title} {\bibinfo {title} {A post-quantum theory
  of classical gravity?},\ }\href {https://arxiv.org/abs/1811.03116} {\bibfield
   {journal} {\bibinfo  {journal} {arXiv preprint arXiv:1811.03116}\ }
  (\bibinfo {year} {2021})}\BibitemShut {NoStop}%
\bibitem [{\citenamefont {Oppenheim}\ and\ \citenamefont
  {Weller-Davies}(2022)}]{Oppenheim2022Constraints}%
  \BibitemOpen
  \bibfield  {author} {\bibinfo {author} {\bibfnamefont {J.}~\bibnamefont
  {Oppenheim}}\ and\ \bibinfo {author} {\bibfnamefont {Z.}~\bibnamefont
  {Weller-Davies}},\ }\bibfield  {title} {\bibinfo {title} {The constraints of
  post-quantum classical gravity},\ }\href
  {https://link.springer.com/article/10.1007/JHEP02(2022)080#article-info}
  {\bibfield  {journal} {\bibinfo  {journal} {Journal of High Energy Physics}\
  }\textbf {\bibinfo {volume} {2022}},\ \bibinfo {pages} {1} (\bibinfo {year}
  {2022})}\BibitemShut {NoStop}%
\bibitem [{\citenamefont {Pearle}(1990)}]{Pearle1990Toward}%
  \BibitemOpen
  \bibfield  {author} {\bibinfo {author} {\bibfnamefont {P.}~\bibnamefont
  {Pearle}},\ }\bibfield  {title} {\bibinfo {title} {Toward a relativistic
  theory of statevector reduction},\ }in\ \href
  {https://doi.org/10.1007/978-1-4684-8771-8_12} {\emph {\bibinfo {booktitle}
  {Sixty-two years of uncertainty}}}\ (\bibinfo  {publisher} {Springer},\
  \bibinfo {year} {1990})\ pp.\ \bibinfo {pages} {193--214}\BibitemShut
  {NoStop}%
\bibitem [{\citenamefont {Dove}\ and\ \citenamefont
  {Squires}(1996)}]{Dove1996Local}%
  \BibitemOpen
  \bibfield  {author} {\bibinfo {author} {\bibfnamefont {C.}~\bibnamefont
  {Dove}}\ and\ \bibinfo {author} {\bibfnamefont {E.~J.}\ \bibnamefont
  {Squires}},\ }\bibfield  {title} {\bibinfo {title} {A local model of explicit
  wavefunction collapse},\ }\href {https://arxiv.org/abs/quant-ph/9605047}
  {\bibfield  {journal} {\bibinfo  {journal} {arXiv preprint quant-ph/9605047}\
  } (\bibinfo {year} {1996})}\BibitemShut {NoStop}%
\bibitem [{\citenamefont {Dove}(1996)}]{Dove1996Explicit}%
  \BibitemOpen
  \bibfield  {author} {\bibinfo {author} {\bibfnamefont {C.~J.}\ \bibnamefont
  {Dove}},\ }\emph {\bibinfo {title} {Explicit wavefunction collapse and
  quantum measurement}},\ \href {http://etheses.dur.ac.uk/5187/} {Ph.D.
  thesis},\ \bibinfo  {school} {Durham University} (\bibinfo {year}
  {1996})\BibitemShut {NoStop}%
\bibitem [{\citenamefont {Breuer}\ and\ \citenamefont
  {Petruccione}(1998)}]{Breuer1998Relativistic}%
  \BibitemOpen
  \bibfield  {author} {\bibinfo {author} {\bibfnamefont {H.-P.}\ \bibnamefont
  {Breuer}}\ and\ \bibinfo {author} {\bibfnamefont {F.}~\bibnamefont
  {Petruccione}},\ }\bibfield  {title} {\bibinfo {title} {Relativistic
  formulation of quantum-state diffusion},\ }\href
  {https://doi.org/10.1088/0305-4470/31/1/009} {\bibfield  {journal} {\bibinfo
  {journal} {Journal of Physics A: Mathematical and General}\ }\textbf
  {\bibinfo {volume} {31}},\ \bibinfo {pages} {33} (\bibinfo {year}
  {1998})}\BibitemShut {NoStop}%
\bibitem [{\citenamefont {Breuer}\ and\ \citenamefont
  {Petruccione}(1999)}]{Breuer1999stochastic}%
  \BibitemOpen
  \bibfield  {author} {\bibinfo {author} {\bibfnamefont {H.-P.}\ \bibnamefont
  {Breuer}}\ and\ \bibinfo {author} {\bibfnamefont {F.}~\bibnamefont
  {Petruccione}},\ }\bibfield  {title} {\bibinfo {title} {Stochastic unraveling
  of relativistic quantum measurements},\ }in\ \href
  {https://doi.org/10.1007/BFb0104400} {\emph {\bibinfo {booktitle} {Open
  Systems and Measurement in Relativistic Quantum Theory}}}\ (\bibinfo
  {publisher} {Springer},\ \bibinfo {year} {1999})\ pp.\ \bibinfo {pages}
  {81--116}\BibitemShut {NoStop}%
\bibitem [{\citenamefont {Pearle}(1999)}]{Pearle1999Relativistic}%
  \BibitemOpen
  \bibfield  {author} {\bibinfo {author} {\bibfnamefont {P.}~\bibnamefont
  {Pearle}},\ }\bibfield  {title} {\bibinfo {title} {Relativistic collapse
  model with tachyonic features},\ }\href
  {https://doi.org/10.1103/PhysRevA.59.80} {\bibfield  {journal} {\bibinfo
  {journal} {Phys. Rev. A}\ }\textbf {\bibinfo {volume} {59}},\ \bibinfo
  {pages} {80} (\bibinfo {year} {1999})}\BibitemShut {NoStop}%
\bibitem [{\citenamefont {Nicrosini}\ and\ \citenamefont
  {Rimini}(2003)}]{Nicrosini2003Relativistic}%
  \BibitemOpen
  \bibfield  {author} {\bibinfo {author} {\bibfnamefont {O.}~\bibnamefont
  {Nicrosini}}\ and\ \bibinfo {author} {\bibfnamefont {A.}~\bibnamefont
  {Rimini}},\ }\bibfield  {title} {\bibinfo {title} {Relativistic spontaneous
  localization: a proposal},\ }\href {https://doi.org/10.1023/A:1025685801431}
  {\bibfield  {journal} {\bibinfo  {journal} {Foundations of Physics}\ }\textbf
  {\bibinfo {volume} {33}},\ \bibinfo {pages} {1061} (\bibinfo {year}
  {2003})}\BibitemShut {NoStop}%
\bibitem [{\citenamefont {Dowker}\ and\ \citenamefont
  {Henson}(2004)}]{Dowker2004SpontaneousLattice}%
  \BibitemOpen
  \bibfield  {author} {\bibinfo {author} {\bibfnamefont {F.}~\bibnamefont
  {Dowker}}\ and\ \bibinfo {author} {\bibfnamefont {J.}~\bibnamefont
  {Henson}},\ }\bibfield  {title} {\bibinfo {title} {Spontaneous collapse
  models on a lattice},\ }\href
  {https://link.springer.com/article/10.1023/B:JOSS.0000028061.97843.84}
  {\bibfield  {journal} {\bibinfo  {journal} {Journal of Statistical Physics}\
  }\textbf {\bibinfo {volume} {115}},\ \bibinfo {pages} {1327} (\bibinfo {year}
  {2004})}\BibitemShut {NoStop}%
\bibitem [{\citenamefont {Bedingham}(2011)}]{Bedingham2011Relativistic}%
  \BibitemOpen
  \bibfield  {author} {\bibinfo {author} {\bibfnamefont {D.~J.}\ \bibnamefont
  {Bedingham}},\ }\bibfield  {title} {\bibinfo {title} {Relativistic state
  reduction dynamics},\ }\href {https://doi.org/10.1007/s10701-010-9510-7}
  {\bibfield  {journal} {\bibinfo  {journal} {Foundations of Physics}\ }\textbf
  {\bibinfo {volume} {41}},\ \bibinfo {pages} {686} (\bibinfo {year}
  {2011})}\BibitemShut {NoStop}%
\bibitem [{\citenamefont {Bedingham}\ \emph {et~al.}(2014)\citenamefont
  {Bedingham}, \citenamefont {D{\"u}rr}, \citenamefont {Ghirardi},
  \citenamefont {Goldstein}, \citenamefont {Tumulka},\ and\ \citenamefont
  {Zangh{\`\i}}}]{Bedingham2014Matter}%
  \BibitemOpen
  \bibfield  {author} {\bibinfo {author} {\bibfnamefont {D.}~\bibnamefont
  {Bedingham}}, \bibinfo {author} {\bibfnamefont {D.}~\bibnamefont {D{\"u}rr}},
  \bibinfo {author} {\bibfnamefont {G.}~\bibnamefont {Ghirardi}}, \bibinfo
  {author} {\bibfnamefont {S.}~\bibnamefont {Goldstein}}, \bibinfo {author}
  {\bibfnamefont {R.}~\bibnamefont {Tumulka}},\ and\ \bibinfo {author}
  {\bibfnamefont {N.}~\bibnamefont {Zangh{\`\i}}},\ }\bibfield  {title}
  {\bibinfo {title} {Matter density and relativistic models of wave function
  collapse},\ }\href {https://doi.org/10.1007/s10955-013-0814-9} {\bibfield
  {journal} {\bibinfo  {journal} {Journal of Statistical Physics}\ }\textbf
  {\bibinfo {volume} {154}},\ \bibinfo {pages} {623} (\bibinfo {year}
  {2014})}\BibitemShut {NoStop}%
\bibitem [{\citenamefont {Tilloy}(2017)}]{Tilloy2017interacting}%
  \BibitemOpen
  \bibfield  {author} {\bibinfo {author} {\bibfnamefont {A.}~\bibnamefont
  {Tilloy}},\ }\bibfield  {title} {\bibinfo {title} {Interacting quantum field
  theories as relativistic statistical field theories of local beables},\
  }\href {https://arxiv.org/abs/1702.06325} {\bibfield  {journal} {\bibinfo
  {journal} {arXiv preprint arXiv:1702.06325}\ } (\bibinfo {year}
  {2017})}\BibitemShut {NoStop}%
\bibitem [{\citenamefont {Bedingham}\ and\ \citenamefont
  {Pearle}(2019)}]{Bedingham2019csl}%
  \BibitemOpen
  \bibfield  {author} {\bibinfo {author} {\bibfnamefont {D.}~\bibnamefont
  {Bedingham}}\ and\ \bibinfo {author} {\bibfnamefont {P.}~\bibnamefont
  {Pearle}},\ }\bibfield  {title} {\bibinfo {title} {On the csl scalar field
  relativistic collapse model},\ }\href {https://arxiv.org/abs/1906.11510}
  {\bibfield  {journal} {\bibinfo  {journal} {arXiv preprint arXiv:1906.11510}\
  } (\bibinfo {year} {2019})}\BibitemShut {NoStop}%
\bibitem [{\citenamefont {Tumulka}(2021)}]{Tumulka2021relativistic}%
  \BibitemOpen
  \bibfield  {author} {\bibinfo {author} {\bibfnamefont {R.}~\bibnamefont
  {Tumulka}},\ }\bibfield  {title} {\bibinfo {title} {A relativistic grw flash
  process with interaction},\ }in\ \href
  {https://link.springer.com/chapter/10.1007/978-3-030-46777-7_23} {\emph
  {\bibinfo {booktitle} {Do Wave Functions Jump?}}}\ (\bibinfo  {publisher}
  {Springer},\ \bibinfo {year} {2021})\ pp.\ \bibinfo {pages}
  {321--347}\BibitemShut {NoStop}%
\bibitem [{\citenamefont {Jones}\ \emph
  {et~al.}(2021{\natexlab{a}})\citenamefont {Jones}, \citenamefont {Guaita},\
  and\ \citenamefont {Bassi}}]{Jones2021Impossibility}%
  \BibitemOpen
  \bibfield  {author} {\bibinfo {author} {\bibfnamefont {C.}~\bibnamefont
  {Jones}}, \bibinfo {author} {\bibfnamefont {T.}~\bibnamefont {Guaita}},\ and\
  \bibinfo {author} {\bibfnamefont {A.}~\bibnamefont {Bassi}},\ }\bibfield
  {title} {\bibinfo {title} {Impossibility of extending the
  ghirardi-rimini-weber model to relativistic particles},\ }\href
  {https://doi.org/10.1103/PhysRevA.103.042216} {\bibfield  {journal} {\bibinfo
   {journal} {Phys. Rev. A}\ }\textbf {\bibinfo {volume} {103}},\ \bibinfo
  {pages} {042216} (\bibinfo {year} {2021}{\natexlab{a}})}\BibitemShut
  {NoStop}%
\bibitem [{\citenamefont {Jones}\ \emph
  {et~al.}(2021{\natexlab{b}})\citenamefont {Jones}, \citenamefont {Gasbarri},\
  and\ \citenamefont {Bassi}}]{Jones2021MassCoupled}%
  \BibitemOpen
  \bibfield  {author} {\bibinfo {author} {\bibfnamefont {C.}~\bibnamefont
  {Jones}}, \bibinfo {author} {\bibfnamefont {G.}~\bibnamefont {Gasbarri}},\
  and\ \bibinfo {author} {\bibfnamefont {A.}~\bibnamefont {Bassi}},\ }\bibfield
   {title} {\bibinfo {title} {Mass-coupled relativistic spontaneous collapse
  models},\ }\href {https://doi.org/10.1088/1751-8121/abf871} {\bibfield
  {journal} {\bibinfo  {journal} {Journal of Physics A: Mathematical and
  Theoretical}\ }\textbf {\bibinfo {volume} {54}},\ \bibinfo {pages} {295306}
  (\bibinfo {year} {2021}{\natexlab{b}})}\BibitemShut {NoStop}%
\bibitem [{\citenamefont {Jones}(2021)}]{Thesis_Jones2021Special}%
  \BibitemOpen
  \bibfield  {author} {\bibinfo {author} {\bibfnamefont {C.~I.}\ \bibnamefont
  {Jones}},\ }\emph {\bibinfo {title} {Special Relativity and Spontaneous
  Collapse Models}},\ \href {https://arts.units.it/handle/11368/2995656} {Ph.D.
  thesis},\ \bibinfo  {school} {Universit{\`a} degli Studi di Trieste}
  (\bibinfo {year} {2021})\BibitemShut {NoStop}%
\bibitem [{\citenamefont {Aharonov}\ and\ \citenamefont
  {Albert}(1980)}]{Aharonov1980States}%
  \BibitemOpen
  \bibfield  {author} {\bibinfo {author} {\bibfnamefont {Y.}~\bibnamefont
  {Aharonov}}\ and\ \bibinfo {author} {\bibfnamefont {D.~Z.}\ \bibnamefont
  {Albert}},\ }\bibfield  {title} {\bibinfo {title} {States and observables in
  relativistic quantum field theories},\ }\href
  {https://doi.org/10.1103/PhysRevD.21.3316} {\bibfield  {journal} {\bibinfo
  {journal} {Phys. Rev. D}\ }\textbf {\bibinfo {volume} {21}},\ \bibinfo
  {pages} {3316} (\bibinfo {year} {1980})}\BibitemShut {NoStop}%
\bibitem [{\citenamefont {Aharonov}\ and\ \citenamefont
  {Albert}(1981)}]{Aharonov1981RelativisticMeasurement}%
  \BibitemOpen
  \bibfield  {author} {\bibinfo {author} {\bibfnamefont {Y.}~\bibnamefont
  {Aharonov}}\ and\ \bibinfo {author} {\bibfnamefont {D.~Z.}\ \bibnamefont
  {Albert}},\ }\bibfield  {title} {\bibinfo {title} {Can we make sense out of
  the measurement process in relativistic quantum mechanics?},\ }\href
  {https://doi.org/10.1103/PhysRevD.24.359} {\bibfield  {journal} {\bibinfo
  {journal} {Phys. Rev. D}\ }\textbf {\bibinfo {volume} {24}},\ \bibinfo
  {pages} {359} (\bibinfo {year} {1981})}\BibitemShut {NoStop}%
\bibitem [{\citenamefont {Aharonov}\ and\ \citenamefont
  {Albert}(1984{\natexlab{a}})}]{Aharonov1984TimeNonRelativistic}%
  \BibitemOpen
  \bibfield  {author} {\bibinfo {author} {\bibfnamefont {Y.}~\bibnamefont
  {Aharonov}}\ and\ \bibinfo {author} {\bibfnamefont {D.~Z.}\ \bibnamefont
  {Albert}},\ }\bibfield  {title} {\bibinfo {title} {Is the usual notion of
  time evolution adequate for quantum-mechanical systems? i},\ }\href
  {https://doi.org/10.1103/PhysRevD.29.223} {\bibfield  {journal} {\bibinfo
  {journal} {Phys. Rev. D}\ }\textbf {\bibinfo {volume} {29}},\ \bibinfo
  {pages} {223} (\bibinfo {year} {1984}{\natexlab{a}})}\BibitemShut {NoStop}%
\bibitem [{\citenamefont {Aharonov}\ and\ \citenamefont
  {Albert}(1984{\natexlab{b}})}]{Aharonov1984TimeRelativistic}%
  \BibitemOpen
  \bibfield  {author} {\bibinfo {author} {\bibfnamefont {Y.}~\bibnamefont
  {Aharonov}}\ and\ \bibinfo {author} {\bibfnamefont {D.~Z.}\ \bibnamefont
  {Albert}},\ }\bibfield  {title} {\bibinfo {title} {Is the usual notion of
  time evolution adequate for quantum-mechanical systems? ii. relativistic
  considerations},\ }\href {https://doi.org/10.1103/PhysRevD.29.228} {\bibfield
   {journal} {\bibinfo  {journal} {Phys. Rev. D}\ }\textbf {\bibinfo {volume}
  {29}},\ \bibinfo {pages} {228} (\bibinfo {year}
  {1984}{\natexlab{b}})}\BibitemShut {NoStop}%
\bibitem [{\citenamefont {Jacobs}\ and\ \citenamefont
  {Steck}(2006)}]{Jacobs2006straightforward}%
  \BibitemOpen
  \bibfield  {author} {\bibinfo {author} {\bibfnamefont {K.}~\bibnamefont
  {Jacobs}}\ and\ \bibinfo {author} {\bibfnamefont {D.~A.}\ \bibnamefont
  {Steck}},\ }\bibfield  {title} {\bibinfo {title} {A straightforward
  introduction to continuous quantum measurement},\ }\href
  {https://doi.org/https://doi.org/10.1080/00107510601101934} {\bibfield
  {journal} {\bibinfo  {journal} {Contemporary Physics}\ }\textbf {\bibinfo
  {volume} {47}},\ \bibinfo {pages} {279} (\bibinfo {year} {2006})}\BibitemShut
  {NoStop}%
\bibitem [{\citenamefont {Jacobs}(2014)}]{Book_Jacobs2014MeasurementTheory}%
  \BibitemOpen
  \bibfield  {author} {\bibinfo {author} {\bibfnamefont {K.}~\bibnamefont
  {Jacobs}},\ }\href {https://doi.org/10.1017/CBO9781139179027} {\emph
  {\bibinfo {title} {Quantum Measurement Theory and its Applications}}}\
  (\bibinfo  {publisher} {Cambridge University Press},\ \bibinfo {year}
  {2014})\BibitemShut {NoStop}%
\end{thebibliography}
\end{document}